\documentclass[twocolumn]{aastex701}
\usepackage{graphicx}
\usepackage[export]{adjustbox}
\usepackage{comment}
\usepackage[varg]{txfonts}
\usepackage{xcolor}
\usepackage{upgreek}
\usepackage[normalem]{ulem}
\usepackage[T1]{fontenc}

\usepackage{fontawesome}

\usepackage{xcolor}
\usepackage{hyperref}

\newcommand{\athena}{{\tt Athena++}}

\renewcommand{\vector}[1]{\ensuremath{\mathbf{#1}}}
\renewcommand{\div}{\ensuremath{\nabla \cdot\,}}
\newcommand{\asinh}{\ensuremath{{\rm asinh}\,}}

\newcommand{\Msun}{\ensuremath{\, M_\odot}}
\newcommand{\Rsun}{\ensuremath{\, R_\odot}}

\newcommand{\gcmc}{\ensuremath{\,\rm g\, cm^{-3}}}
\newcommand{\cmc}{\ensuremath{\,\rm cm^{-3}}}

\newcommand{\kms}{\ensuremath{\,\rm km\,s^{-1}}}

\newcommand{\cs}{\ensuremath{c_{\rm s}}}

\newcommand{\A}{{\cal A}}
\newcommand{\Mach}{\ensuremath{{\cal M}}}
\newcommand{\srm}[1]{\ensuremath{_{\rm{#1}}}}

\newcommand{\Mx}{\ensuremath{{\rm Mx}}} 

\newcommand{\simD}{{\tt S1}}
\newcommand{\simDt}{{\tt S2}}
\newcommand{\simp}{{\tt BH1}}
\newcommand{\simo}{{\tt BH1L}}
\newcommand{\simH}{{\tt BH2}}
\newcommand{\simHZ}{{\tt BHz}}
\newcommand{\simHL}{{\tt BH2L}}

\makeatletter
\newcommand{\github}[1]{%
   \href{#1}{\faGithubSquare}%
}
\makeatother

\DeclareRobustCommand{\uppartial}{\text{\rotatebox[origin=t]{20}{\scalebox{0.95}[1]{$\partial$}}}\hspace{-1pt}}

\newcommand{\pardir}[2]{\ensuremath{\frac{\uppartial #2}{\uppartial #1} }}

\renewcommand{\i}{\ensuremath{{\rm i}}}
\newcommand{\e}{\ensuremath{{\rm e}}}
\newcommand{\diff}{\ensuremath{{\rm d}}}

\begin{document}

\title{Tidal disruption of a magnetized star}

\author[orcid=0000-0002-6429-0436, sname='Abolmasov']{Pavel Abolmasov}
\affiliation{The Raymond and Beverly Sackler School of Physics and Astronomy, Tel Aviv University, Tel Aviv 69978, Israel}
\email[show]{pavel.abolmasov@gmail.com}  

\author[orcid=0000-0002-6429-0436, sname='Bromberg']{Omer Bromberg}
\affiliation{The Raymond and Beverly Sackler School of Physics and Astronomy, Tel Aviv University, Tel Aviv 69978, Israel}
\email[]{omerbr@tauex.tau.ac.il}  

\author[orcid=0000-0001-7572-4060, sname='Levinson']{Amir Levinson}
\affiliation{The Raymond and Beverly Sackler School of Physics and Astronomy, Tel Aviv University, Tel Aviv 69978, Israel}
\email[]{levinson@tauex.tau.ac.il}  

\author[orcid=0000-0002-4534-7089, sname='Nakar']{Ehud Nakar}
\affiliation{The Raymond and Beverly Sackler School of Physics and Astronomy, Tel Aviv University, Tel Aviv 69978, Israel}
\email[]{udini@tauex.tau.ac.il}  


\begin{abstract}
Tidal disruptions of stars by supermassive black holes in galactic centers (TDEs) are now being actively studied both theoretically and observationally. They are observed throughout the electromagnetic spectrum, from radio to gamma-rays. It is still unclear how the emission is produced and, in particular, what is the role of the magnetic field of the disrupted star. There are many ways how magnetic fields might affect the dynamics of a TDE. They are likely responsible for the angular momentum transfer in the accretion disk formed at later stages and thus affect the radiation associated with the disk. Magnetic fields are also an important requirement for the formation of relativistic jets, that are seen in some TDEs.
 The goal of our study is to connect the field within the star to the fields that develop during the fallback and disk accretion. 
 Using the fluid-dynamic code \athena, we perform a large-scale three-dimensional adaptive-mesh magnetohydrodynamic simulation of a tidal disruption of a magnetized star. 
 The fallback stream returning to the black-hole vicinity after the disruption contains smooth magnetic fields aligned with the stream lines. Formation of a nozzle shock near the pericenter of the initial orbit leads to a turbulent eccentric disk-like structure where the field is amplified and entangled on the local dynamic time scales up to approximate equipartition.
 The resulting field is mildly anisotropic and has a typical length several times smaller than the pericenter distance. The properties of the field are consistent with the early stages of turbulent dynamo. 
 \end{abstract}



\section{Introduction} 

There is a general consensus that the centers of galaxies contain a population of supermassive black holes (BHs). 
Their properties and evolutionary paths are tightly connected to those of the host galaxies \citep{graham_review}.
The properties of the central BHs may be studied for a number of nearby objects (especially at the center of our Galaxy; see \citet{Genzel10} for a review) or in the exceptional cases when the galactic nucleus undergoes rapid accretion, which makes the BH visible through the release of gravitational energy as an Active Galactic Nucleus (AGN). 
A unique opportunity to gain substantial information on quiescent BHs in distant galaxies is provided by the tidal disruptions of individual stars passing close enough to the BH on nearly parabolic orbits. 
This phenomenon is known as a Tidal Disruption Event (TDE). 

TDEs were predicted theoretically in a number of papers \citep{hills75,Rees88,EK89}. 
A star flying by in an eccentric orbit is subject to the tidal forces from the BH. 
At a sufficiently small distance, they overcome the self-gravity of the star, and the stellar material spreads to a fan of orbits with different energies and eccentricities. 
The bound part of the disrupted material returns in the form of a fallback stream and eventually accretes onto the BH.
The fallback mass accretion rate easily overcomes the Eddington limit for the BH. 
The radiative inefficiency of the super-Eddington regime of accretion and its tendency to launch winds \citep{1973A&A....24..337S, poutanen07} suggests that much more than half of the stellar mass is ultimately ejected during the accretion process. 

The earliest TDE candidates were reported in the 1990s and were found in the data of the ROSAT all-sky survey in soft X-rays \citep{bade96,Komossa99}. 
Since then, numerous TDEs have been discovered through X-ray, UV, and optical observations. A detailed observational review is given by \citet{Gezari}. 
Though their optical/UV as well as the X-ray spectra appear to be thermal, the inferred temperatures differ by two orders of magnitudes, indicating a different origin for the two spectral ranges.
The temperature of the X-ray emission falls in the range of $0.03-0.1$~keV, expected for the inner disk of an $\sim 10^6\Msun$  BH accreting close to its Eddington limit \citep{saxton20}. 
This is consistent with the fact that the X-ray emission is usually delayed with respect to the optical, as the stellar material requires a considerable amount of time to return after the initial disruption and to form a disk. 
The optical and UV emission, on the other hand, are characterized by smooth light curves with a power-law decline consistent with the theoretical expectations for the mass fallback rate \citep{Phinney89}. 
Possible sources of this emission are shock waves within the fallback stream and reprocessing of soft X-ray or EUV photons, that may be impossible to observe directly because of absorption by the accretion streams.

A considerable fraction of TDEs (about one half, see \citealt{2024ApJ...971..185C, 2025ApJS..278...36G}) shows delayed radio emission probably connected to sub-relativistic ($0.01-0.1c$) ejecta launched during the disruption process. 
The emission is consistent with the synchrotron radiation of a system of shock waves launched by ejection of a small fraction of stellar material in the circumnuclear medium with number densities $\sim 10-10^4\cmc$ and magnetic fields $0.1-1$G. 

A small minority of events show strong and rapidly decaying radio emission \citep{2020SSRv..216...81A} that might be associated with relativistic jets. 
This subclass is also distinguished by early and rapidly evolving X-ray and gamma-ray radiation \citep{2011Sci...333..199L, 2011Sci...333..203B, 2015MNRAS.452.4297B, 2012ApJ...753...77C, 2024ApJ...974..149E} consistent with a relativistic outflow forming at about the time of the optical maximum.
In the case of three of the four known jetted TDEs, \object{Swift~1644+57} \citep{2013ApJ...767..152Z}, \object{Swift~2058+05} \citep{2015ApJ...805...68P}, and \object{AT2022cmc} \citep{2024ApJ...974..149E}, the X-ray and gamma radiation associated with the jet is rapidly quenched after about a year. 
Early existence of a jet is an indirect evidence for the accumulation of large magnetic flux near the BH horizon, and thus for considerable magnetization of the fallback material. 
The origin of this magnetic field is unclear, as the expected magnetization of the fallback matter is unlikely to significantly exceed that of the disrupted star \citep{bonnerot16}.
It is likely (see discussion in \citealt{2024ApJ...974..149E}) that the jetted TDEs contain less massive BHs than the bulk of the TDE population, which makes it easier to produce powerful relativistic jets during the early stages of the accretion process, when the mass fallback rate is highly super-Eddington. 
All the scenarios involving early jet formation require substantial amplification of the stellar magnetic field during and/or immediately after the disruption of the star.

The delayed and long-lasting radio emission of well-studied TDEs hints upon the existence of a population of galactic nuclei where past tidal disruptions are responsible for present radio emission. 
Such a connection was proposed for the population of the so-called compact symmetric objects (CSO2s, see \citealt{2021A&ARv..29....3O} for review) potentially powered by TDEs \citep{readhead2024,sullivan2024}.
This interpretation favors TDEs of giant branch stars, which can naturally explain the time scales and event rates of CSO2s, although shorter TDEs of main-sequence stars are not ruled out.  

Another potential evidence for magnetization is the high (about 25 per cent) linear polarization in the optical range reported for \object{AT2020mot} by \citet{2023Sci...380..656L}. 
Such a polarization degree is difficult to produce by scattering, unless the primary source of radiation is obscured and the observed emission is scattered at a favorable angle. 
Alternatively the high observed polarization can originate from synchrotron emission, which in this case points to strong magnetic fields with a relatively ordered configuration.

\subsection{Simulating TDEs}

Tidal disruption is a complex, multi-stage process, involving (i) disruption per se during the initial pericenter passage of the star, (ii) a nearly ballistic evolution of the stellar debris cloud, part of which falls back towards the black hole as a fallback stream, and (iii) an accretion phase, during which the eccentric fallback stream evolves into an accretion disk via self-interaction. 
This division was suggested by \citet{lodato20} in a review focused on numerical simulations of TDEs, as the different phases set different challenges and require different approaches.

During the initial linear stretching phase, the
size of a main-sequence star is typically tens (for a BH mass $10^3- 10^6\Msun$) to hundreds (for a BH mass $\gtrsim 10^6\Msun$) of times smaller than the tidal disruption radius. 
A simulation box containing both the black hole and the star should be initially filled by about $10^{-6}$ of its volume.
The ambient circumnuclear matter has a density typical for the interstellar medium (otherwise, accretion of the background material onto the BH will make it an AGN), meaning that a realistic dynamic range for the density in the problem is about twenty orders of magnitude.
Another complication making grid-based simulations of TDEs challenging is the supersonic accelerated motion of the star relative to the grid.

The dimensions of the fallback stream are even more extreme. 
The debris of the disrupted star are spread along highly eccentric orbits with the typical apocenter distances of thousands of stellar radii and more. 
Also, their relative velocities are now highly supersonic. 
The transverse sizes of the stream may be as small as the initial size of the star. 
During the final stage, an accretion disk is formed at distances of about the tidal disruption radius. 
As a consequence of angular momentum transfer, its outer edge is moving outwards with time, that should be taken into account while choosing the size of the simulation box. 

To avoid simulating all the empty space, it is reasonable to use Lagrangian approach. 
One such scheme is the Smooth Particle Hydrodynamics (SPH, \citealt{lucy77}).
In fact, tidal disruption of a star was one of the first problems for which SPH was used \citep{NK82}.
The SPH approach treats the star as a group of interacting particles. 
Obviously, this allows to avoid simulating the empty space outside the star.
The most critical downsides of the SPH approach are related to the treatment of discontinuities such as shock waves \citep{price08}. 
In addition, many hydrodynamic effects such as turbulence and dissipation may be reproduced with insufficient accuracy. 
For this reason, studying TDEs with grid-based simulations is important, as they are capable of grasping  phenomena such as fluid instabilities or shock dynamics, that are likely to be lost in SPH simulations. 

An alternative approach that allows one to simulate efficiently the early stages of the disruption process is using a grid-based code and a reference frame moving with the star. 
In such a setup, it is possible to use a much smaller simulation box and to avoid the numerical errors related to moving the star across the grid. 
The approach works well during the early stages of the TDE, but after the first pericenter passage the size of the box should be adjusted to account for the changing size of the debris cloud. In addition, with time, the relative velocities of the gas streams become comparable to the virial velocity, making the star frame equally poorly fitted for studying their dynamics.

As a result, most numerical studies of TDEs involve two steps. 
The disruption stage is simulated using either SPH or a star-centered grid-based setup. 
The advanced stages are considered on a larger grid in the BH reference frame. 

As we have seen earlier, magnetic fields are an important factor at the late stages, during the formation of the accretion disk and relativistic jets. 
Many simulations focus on the late evolution of a circularized accretion flow fed by a fallback stream.
For example, \citet{dai18} perform elaborate numerical simulations of the super-Eddington thick disk evolution thought to be responsible for the X-ray and EUV radiation of TDEs near the peak of the fallback rate. 
The initial mass distribution and magnetic fields in this case have to be set by hand, and the accreting mass is injected into the simulation domain as a narrow magnetized fallback stream \citep{sadowski16, kaur23, 2025MNRAS.540.1215C}.
In reality, the properties of the magnetic fields within the stellar debris cloud are determined by the seed magnetic fields inside the star and the evolution of the field during the disruption process and the fallback. 
This makes it extremely important to track the evolution of a magnetized star disrupted by tidal forces consistently, through all the stages of the disruption process.

There are only few numerical studies of tidal disruptions of magnetized stars, mostly limited to the initial stage up to the formation of tidal tails. 
One is by \citet{Guillochon}, who used a grid-based approach with adaptive mesh refinement (AMR) in a frame co-moving with the star.
Even this early evolution shows some of the general features that are crucial for the development of the accretion flow, such as orientation of the magnetic field lines along the tidal streams and increasing magnetization. 
Another important study was \citet{bonnerot17} who used modified SPH techniques to study the evolution of the magnetic field in the tidal streams. 
The results presented in these papers covers only the first stage of the disruption, after which the magnetized tidal streams experience further shear, compression and stretching followed by a violent and poorly known circularization process.  

The role of the magnetic fields in the disk is crucial in two ways: first, they are responsible (at least to a certain degree) for angular momentum transfer after circularization, and thus for the late-stage light curve of the event; second, the formation of a relativistic jet and thus the radio emission observed in some sources require the presence of a magnetic field in the disk. 
No studies so far have traced the magnetic field evolution all the way from the stellar interior to disk formation. 

\subsection{Fallback stream circularization}

The transition between a nearly parabolic fallback stream and a nearly circular disk is a controversial issue in TDE studies. 
In the first place, a geometrically thin fluid stream of matter following a Keplerian orbit dissipates very little energy and does not naturally form a disk. 
Several dissipation mechanisms were considered that can efficiently circularize the flow, including relativistic apsidal precession leading to self-intersection of the orbit \citep{Rees88, 2015ApJ...804...85S} and \emph{nozzle shocks} near the pericenter of the orbit \citep{kochanek94}. 
It is unclear which process drives the circularization and how fast the accretion disk is formed. 
Its is also unclear if the disk is relevant for the observed optical/UV emission, probably generated by shocks during the circularization process \citep{2023ApJ...957...12R}. 
Other factors that may affect the circularization process are hydrodynamic instabilities in the flow, magnetic stresses, and interaction with a pre-existing accretion disk around the BH. 
The latter becomes relevant if some part of stellar material has circularized before the bulk of the debris. 
This is the case of \emph{runaway circularization} proposed by \citet{steinbergstone} and supported by some numerical results. 

The fallback stream in its major part has total net mechanical energy close to zero and net angular momentum much smaller than the local Keplerian value.
The low angular momentum of the fallback stream has lead to a suggestion \citep{CB14,metzger22} that instead of circularization, the flow retains most of its heat and forms a quasi-spherical optically thick flow cooled by the radiation from its surface. 

\subsection{Outline of this paper}

In this paper, we consider the consequences of a tidal disruption of a star containing weak magnetic fields. 
The goal is to trace the evolution of the magnetic field from the seed fields within the star all the way to the accretion disk. 
To solve the problem, we perform magnetohydrodynamic (MHD) simulations of the star affected by external tidal forces and the tidal streams formed after the disruption, switching from the stellar reference frame to the frame of the BH and using AMR to increase the efficiency of the calculations. 
Our numerical setup is described in Section~\ref{sec:num}. 
The results are presented in Section~\ref{sec:res} and discussed in Section~\ref{sec:disc}. 
We conclude in Section~\ref{sec:conc}. 

\section{Physical problem and numerical setup}\label{sec:num}

\subsection{Physical scaling and units}

The tidal disruption problem involves spatial scales differing by several orders of magnitude. 
First, one needs to set the mass $M_*$ and radius $R_*$ of the star and its internal structure. 
Both quantities were assumed to be about the solar values, $M_*\sim 1-3\Msun$, and $R_*\sim 5-10\Rsun$. 
The BH mass is expected to be several orders of magnitude larger than the mass of the star. 
Most TDE observations argue for $M_{\rm BH} \sim 10^6\Msun$, though lower $M_{\rm BH} \lesssim 10^5\Msun$ are of special interest as potentially associated with jetted TDEs. 
Also, lower BH masses are easier to simulate numerically. 
We will consider a small BH mass of $M_{\rm BH} \sim 3\times 10^3\Msun$, still considerably larger than the mass of the star. 
The masses and the stellar radius imply a tidal disruption radius of
\begin{equation}
    r_{\rm t} = R_* \left( \frac{M_{\rm BH}}{M_*} \right)^{1/3} \sim 100\Rsun.
\end{equation}
After the disruption, stellar material is set onto a range of orbits with different net energies, eccentricities, and semi-major axes.  
The most bound of the orbits has the semi-major axis (or the apocenter distance) of about
\begin{equation}\label{E:amin}
    a_{\rm min} \sim \frac{r_{\rm t}^2}{R_*} \sim R_* \left( \frac{M_{\rm BH}}{M_*}. \right)^{2/3} \sim 10^3\Rsun.
\end{equation}
Hereafter, capital $R$ and lowercase $r$ would be used for the distances towards the center of the star and towards the BH center, respectively. 
In order to simulate the entire process from the stellar disruption to the formation of a disk, we need $a\srm{min}$ to be located well within the computational box. 
This makes simulations with smaller black hole masses much more computationally efficient and easy to perform. 
The star is initially set on a parabolic orbit with a pericenter radius $r\srm{p} \lesssim r\srm{t}$. 
It is convenient to describe the orbit by the so-called penetration parameter
\begin{equation}
    \beta = \frac{r\srm{t}}{r\srm{p}}.
\end{equation}
In our simulations, we set $r_{\rm p} = 100\Rsun$, implying $\beta \sim 2-3$. 
The moderately deep penetration ensures that during the first fly-by the star will be totally disrupted but would not suffer additional compression shocks that should form when $\beta \gtrsim 5$ as described by \citet{LC86}.

The initial sound velocity inside the star is $\cs \sim \sqrt{GM_*/R_*}$.
Thus, the Mach number of its orbital motion in the BH frame is $\Mach \sim \sqrt{\beta} \left( M_{\rm BH}/M_*\right)^{1/3}$ near the pericenter of the orbit. 
Heating by shock waves rises the speed of sound by a factor of $\sim \beta $ \citep{CL83}, and decreases the Mach number to $\displaystyle \Mach \sim \frac{1}{\sqrt{\beta}} \left( M_{\rm BH}/M_*\right)^{1/3}\gtrsim 10$. 
As high Mach numbers may be challenging for grid-based simulations, during the early stages it makes sense to choose a reference frame moving with the star. 
Later on, as the relative velocities of the tidal streams become comparable to the virial velocities, the advantage of the co-moving frame disappears.
As one may see, larger mass ratios lead to larger Mach numbers and are thus more challenging from the computational perspective. 

The TDE problem has at least two important dynamical time scales. 
The typical time scale for the evolution of the system at the tidal radius set by the free-fall time at $r\srm{t}$
\begin{equation}
\begin{array}{l}
 \displaystyle   t^* = t_{\rm ff}(r_{\rm t}) \simeq 2\uppi \sqrt{\frac{r_{\rm t}^3}{GM_{\rm BH}}} = 2\uppi \sqrt{\frac{R_{\rm *}^3}{GM_{\rm *}}}\\
   \displaystyle \qquad{} \simeq 50\left(\frac{R_*}{10\Rsun}\right)^{3/2}\left(\frac{M_*}{3\Msun}\right)^{-1/2} {\rm h},\\
    \end{array}
\end{equation}
which is also of the order of the dynamical time of the star. The fly-by duration is about $\beta^{-3/2}t^*$.
The second important time is the fallback time from the apocenter
\begin{equation}
    t_{\rm fb} = \uppi \sqrt{\frac{a_{\rm min}^3}{GM_{\rm BH}}} = \frac{1}{2} \sqrt{\frac{M_{\rm BH}}{M_*}} t^*.
\end{equation}
That is, for our simulated parameters, of the order months to years. 
This is comparable to the typical duration of the optical/UV maximum of a TDE, and the delay of its X-ray light curve found in observational data, which are usually interpreted as a manifestation of $t_{\rm fb}$. 
The likely reason that the fallback time in our simulation is comparable to the observed ones (if the interpretation of the observed timescales is correct), although our simulated BH mass is significantly lower than the realistic BH masses, is that our simulated stellar radius is larger by about an order of magnitude than the radii of typical disrupted stars.

\subsection{Simulation setup}\label{sec:num:sim}

All our simulations were run with the MHD code \athena~\citep{athena, athenapp}.
To address the multi-scale nature of the problem, the simulation is run in several stages, first using a grid moving with the star, and then a grid where the BH is static. 
Both use cubic Cartesian meshes with several levels of AMR and solve the equations of ideal MHD. 
As \athena\ does not support mesh extension within the code, we have added a module that reads data from HDF5 outputs and restores the hydrodynamic quantities and magnetic fields on a new mesh\footnote{This branch of the code is available at \url{https://github.com/pabolmasov/athena-remap.git}\github{https://github.com/pabolmasov/athena-remap.git}}. 
The properties of the grids used in our simulations are given in Table~\ref{tab:mod}. 

The physics included in the stellar- and BH-frame runs is different, we summarize the specific properties of the two approaches in the following subsections.
The equation of state for all the simulations is adiabatic with $\gamma = 5/3$. 
Both setups include the gravity of the BH as a pseudo-Newtonian potential modified near the BH to avoid singularity. 
Gravity is a mass force directed towards the BH and equal by the absolute value to 
\begin{equation}\label{E:force}
  \displaystyle  g(r) = \left\{ 
\begin{array}{cc}
\displaystyle     \frac{GM_{\rm BH}}{r^2} \left(\frac{1-\frac{2GM}{r_{\rm 0}c^2}}{1-\frac{2GM}{rc^2}}\right)^2 &  r>r_{\rm 0}\\
\displaystyle      \frac{GM_{\rm BH}}{r_{\rm 0}^2} \frac{r}{r_{\rm 0}}&  r<r_{\rm 0},\\
\end{array}
    \right.
\end{equation}
where $r_{\rm 0} = 10\Rsun \gg GM_{\rm BH} /c^2$, except one specific run with $r_0 = 1\Rsun$ (\simHZ, see Table~\ref{tab:mod}). 
The outer part ($r>r_{\rm 0}$) corresponds to the Paczy\'nski-Wiita potential \citep{PW80}, while the inner region with a linear gravity law allows  to avoid singularities during calculations.
For the considered BH mass, GR effects are negligible. 
In particular, the last stable orbit radius for $M_{\rm BH} = 3\times 10^3\Msun$ is $R_{\rm ISCO} \simeq 0.04\Rsun$, that is three orders of magnitude smaller than the expected circularization radius and well below the resolution limit. 
In all the runs, we use outflow boundary conditions for hydrodynamic variables as well as for the magnetic fields. 

\subsection{Stellar-frame stage}\label{sec:num:star}

The stellar-frame run includes self-gravity and starts with a spherical star moving along a parabolic trajectory as a rigid body without rotation. 
The coordinates of the star $\vector{r}_*(t)$ as a function of time are calculated using the formulae of Appendix~\ref{sec:app:parabolic}.
The pericenter of the orbit is located at $r_{\rm p} = 100\Rsun$, and the initial radius is $R_0 \simeq 435\Rsun$. 
The star is assumed to be initially in the state of hydrostatic equilibrium. 
For the mass distribution within the star, we use an $n=5$ polytrope solution. 
Such a configuration has the advantage of being calculable analytically, but it has an infinitely extended outer corona with $\rho \propto R^{-5/2}$, which we cut artificially with an approximately exponential function at a cut-off radius $R_{\rm CO}=13.56\Rsun$. 
A detailed description of the density profile is given in Appendix~\ref{sec:app:Emden}. 
The parameters used where $\alpha = 6\Rsun$, $R_{\rm CO} = 13.56\Rsun$, $\Delta R_{\rm CO} = R_{\rm CO} / 4 \simeq 3.39\Rsun$. 
The total mass of the polytropic configuration is $3\Msun$, that is reduced by the cut-off radius to $M_* \simeq 1.8\Msun$.
The corresponding tidal radius value is $r_{\rm t} \simeq 120\Rsun$. 
Depending on the exact definition of the stellar mass and radius, the tidal radius position is in the range $100-200\Rsun$ corresponding to $\beta \sim 1-2$. 
We use a tracer quantity $s$ initially equal to $1$ for $R<R_*$ and $0$ outside.
The radius $R_*$ encloses about $M(R<R_*) \simeq 1\Msun$, or 60 per cent of the stellar mass.

The star is immersed in a uniform ambient medium with density $\rho_{\rm bgd} = 10^{-10}\Msun \Rsun^{-3}$ and pressure $P_{\rm bgd} = 10^{-9} G\Msun^2\Rsun^{-4}$.
The maximal density and pressure within the star are, respectively, $\rho_{\rm c} \simeq 8.5\times 10^{-4}\Msun \Rsun^{-3}$ and $P_{\rm c} \sim 3\times 10^{-5} G\Msun^2 \Rsun^{-4}$. 
The density contrast is thus $\lesssim 10^6$, that is considerably smaller than the real one. 
Moreover, the expansion of the stellar debris reduces it even further. 
To address this issue, we applied a mass boost procedure when switching to the BH frame(see Section~\ref{sec:num:BH}). 
The dynamic effect of the ambient medium is mostly related to the mass swept by the stellar debris, that we estimate in Appendix~\ref{sec:app:dyne}. 

The initial simulation box has the dimensions of $500\times 500\times 125\Rsun^3$ (smaller in the vertical direction). 
The basic mesh is $256\times 256\times 64$ cells, corresponding to a non-refined spatial resolution of about $2\Rsun$ (see Table~\ref{tab:mod}). 
We use four levels of refinement, yielding a resolution of about $0.2\Rsun$ within the star. 
The condition for refinement is a combination of the tracer value $s > 0.5$, density $\rho > 10\rho_{\rm bgd}$, and a critical cell-to-cell density variation $\Delta \rho \geq 0.01 \rho$. 

The star is initially magnetized in the range of radii $R = 2-8\Rsun$. 
The problem generator allows for two different magnetic field configurations, dipolar and toroidal, described in detail in Appendix~\ref{sec:app:fields}.
Here, we show only the results for the dipolar topology with the magnetic axis perpendicular to the orbital plane.
The initial magnetic field strength is normalized  such that the initial plasma $\beta_{\rm m}=  8\uppi P / B^2$ has the minimal initial value of $10^3$.

\subsection{Remapping}\label{sec:num:remap}

To prevent the debris from expanding beyond the boundaries of the computational domain, we remap the simulation onto a larger grid, modifying both the mesh size and resolution. 
To this end we have developed a remapping technique that interpolates the hydrodynamical quantities and magnetic fields while retaining zero divergence of the latter. 
While hydrodynamic quantities may be linearly interpolated from the original mesh to the new one, restoring the magnetic fields on the new grid is less straightforward. 
Interpolation of the magnetic field components involves calculation of vector potentials on an intermediate regular grid (see Appendix~\ref{sec:app:avec}). 

\begin{figure*}
\includegraphics[width=1.0\textwidth]{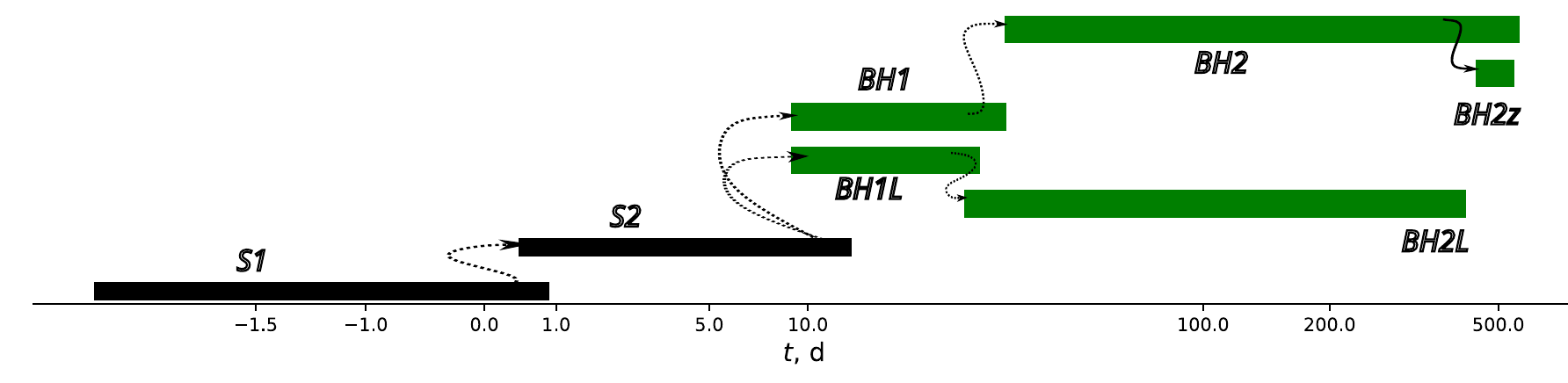}
 \caption{Time spans and remapping sequences in the simulation. Dotted lines show remappings, solid line is used to show a restart with an increased resolution. Black bars show stellar-frame self-gravitating models, BH-frame models are shown with broader green bars.}\label{fig:modelgantt}
\end{figure*}

\begin{figure*}
\includegraphics[width=1.0\textwidth]{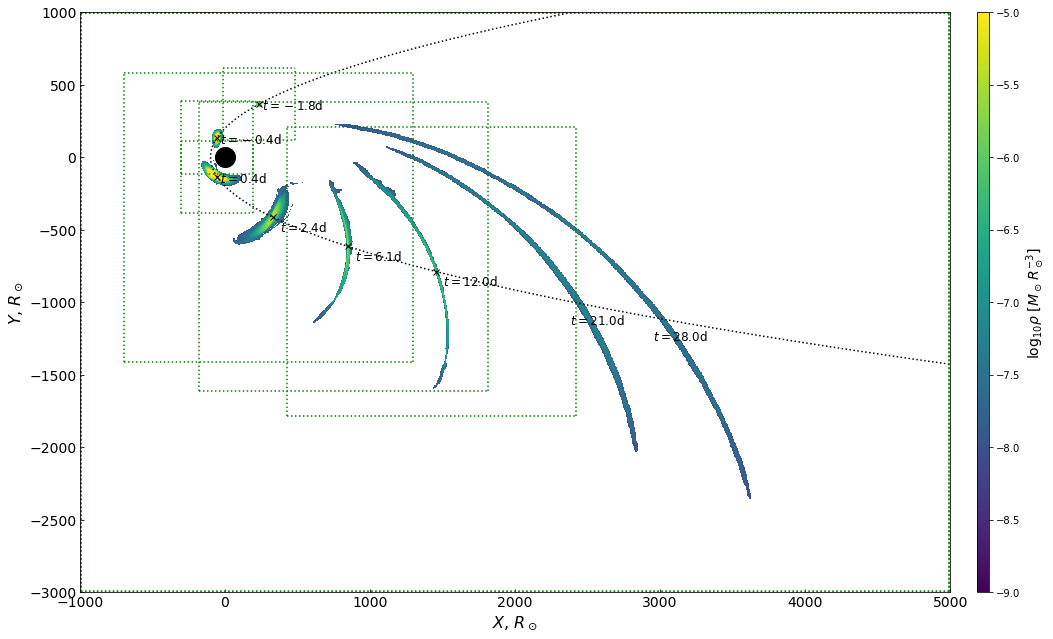}
 \caption{Multi-exposure plot showing the evolution of the stellar debris from the initial hydrostatic configuration towards the fallback. Each image is a snapshot of the density (log-scale) in the equatorial plane, cut off at the level of $10^{-8} \Msun \Rsun^{-3}$. First three frames are from the stage \simD, three following from \simDt, and the last one from \simp. Green dotted rectangles are the boundaries of the simulation boxes. The boundaries of the plot coincide with the boundaries of the BH-frame simulation domain.}\label{fig:array}
\end{figure*}

The first remapping was applied to the star-frame problem when the size of the debris cloud becomes comparable to the size of the simulation box. 
The simulation is then restarted in a box with the same resolution but several times larger in the $X$ and $Y$ directions (in the orbital plane). The remapping sequence for all the model stages is shown in Fig.~\ref{fig:modelgantt} and Table~\ref{tab:mod}.
In Fig.~\ref{fig:array}, we show several snapshots of the density distribution in the equatorial plane during the early stages of evolution, from the initial conditions towards the formation of the fallback stream. 
We also show the sizes of the simulation boxes, that change from $500\times 500\Rsun$ for the initial mesh to $6000\times 4000\Rsun$ for the (smaller) black-hole-frame mesh used at the later stages.

\subsection{BH-frame stage}\label{sec:num:BH}

About 10d after the pericenter passage, when the bound part of the debris cloud starts falling towards the BH, we performed another remapping, switching from the reference frame moving with the star to the BH reference frame.
At this point we switch off self-gravity, as we expect its impact on the supersonic ballistic tidal stream to be negligible. 
However, as the pressure inside the stream is not uniform and its gradients are partly supported by self-gravity, simply turning off gravity would have created an episode of unphysical accelerated expansion (see Section~\ref{sec:res:shape} for a detailed discussion on the stream shape and the role of self-gravity during the expansion stage).
Another issue that should be addressed at this stage is the drag of the ambient medium. 
In Appendix~\ref{sec:app:dyne}, we show that the amount of ambient matter that the debris cloud is going to interact with at the later stages of the simulation is of the order of the mass of the star. 

To address these challenges, we took two additional steps. 
First, we applied a \emph{mass boost} during the remapping stage by multiplying the density by a factor depending upon the tracer as
\begin{equation}
    \rho_1  / \rho_0 = 1 + s(b-1), 
\end{equation}
where $b=10^3$ is a constant boost factor. 
Such a scaling ensures that stellar material distribution is retained. 
In the absence of self-gravity, the increase in mass does not affect the physics. 
Magnetic field is boosted by the square root of this factor. 
We did not apply the boost factor to pressure. 
Instead, we overwrite the pressure distribution with a hydrostatic isothermal configuration
\begin{equation}
    P = P_{\rm bgd} \exp\left( \frac{GM}{c^2_{\rm s, \, bgd}} \left[ \frac{1}{r} -  \frac{1}{r_*}\right]\right),
\end{equation}
where $r$ is the distance towards the BH center, $r_*$ is the radial coordinate of the calculated stellar center of mass (see Appendix~\ref{sec:app:parabolic}), and $c_{\rm s, \, bgd}^2 = \gamma P_{\rm bgd} / \rho_{\rm bgd}$ is the speed of sound of the ambient matter. 
The density was ascribed to the background density $\rho_{\rm bgd} = \gamma P / c^2_{\rm c, \, bgd}$ or the (boosted) density interpolated during the remapping procedure, whatever is higher. 
As a result, the entire box is in an approximate pressure equilibrium, the hot ambient gas confines the stellar debris by its pressure, but does not affect the dynamics by its inertia. 

\begin{table*}\centering
\caption{Simulation stages.}\label{tab:mod}
\begin{tabular}{lccccccc}
\hline \hline
ID & $t$ range [d] & frame & limits $[\Rsun]$ & resolution $[\Rsun]$ & self-gravity & mass boost & comment\\
&&&$x\times y\times z$&&&&\\
\hline
\simD & $-1.8..1.0$ & star  & $500\times 500 \times 125$&  $0.2..2$ &\checkmark & $1$ & disruption stage\\
\hline
\simDt & $0.5..12$ & star  & $1600\times 1600 \times 200$ & $0.8..6$ &\checkmark & $1$ & debris expansion\\
\hline
\simp & $10..30$ & BH  & \begin{tabular}{l}
     $-1500..5000$  \\
   $\times -2500..1000$ \\
   $\times  -250..250$\\
\end{tabular} & $1..16$ & $-$ & $10^3$ & fallback formation\\
\hline
\simo & $10..35$ & BH  & \begin{tabular}{l}
     $-2500..7500$  \\
   $\times -3000..2000$ \\
   $\times  -250..250$\\
\end{tabular} & $1..16$ & $-$ & $10^3$ & fallback formation\\
\hline
\simH & $30..520$ & BH  &\begin{tabular}{l}
     $-1500..5000$  \\
   $\times -2500..1500$ \\
   $\times  -600..600$\\
\end{tabular} & $2..16$  & $-$ & $10^3$ & 
\begin{tabular}{l}
fallback ($t\sim 40-250$d)\\
abalone ($t\sim 60-300$d)\\
circularization ($t\sim 300-500$d)\\
\end{tabular}\\
\hline
\simHL & $30..500$ & BH  &\begin{tabular}{l}
     $-2500..7500$  \\
   $\times -3000..2000$ \\
   $\times  -500..500$\\
\end{tabular} & $2..16$  & $-$ & $10^3$ & 
\begin{tabular}{l}
fallback ($t\sim 40-250$d)\\
abalone ($t\sim 60-300$d)\\
circularization ($t\sim 300-500$d)\\
\end{tabular}\\
\hline
\simHZ & $480..500$ & BH  & \begin{tabular}{l}
     $-1500..5000$  \\
   $\times -2500..1500$ \\
   $\times  -600..600$\\
\end{tabular}   & $1.2..18$ & $-$ & $10^3$ & disk zoom-in, $r_0 = 1\Rsun$\\
\hline
\end{tabular}
\end{table*}

\section{Results}\label{sec:res}

\subsection{Debris cloud evolution}\label{sec:res:shape}

Distortion of the star shape involves stretching along the orbit and contraction along the vertical direction, followed by rapid expansion, resulting in a flattened pancake-shaped structure perpendicular to the orbital plane. 
It is possible to describe the shape of the stellar debris cloud in terms of a covariance matrix in the stellar mass distribution
\begin{equation}
    C_{ij} = \langle R_i R_j\rangle - \langle R_i\rangle \langle R_j\rangle,
\end{equation}
where $R_i$ are the radius vector components measured from the center of the star, $i, j = 1-3$, and the angle brackets denote tracer-weighed spatial averaging, 
\begin{equation}
    \langle x\rangle = \frac{\int x s \diff V}{\int s \diff V}.
\end{equation}
The shape of the debris cloud, as long as it is reasonably well approximated by an ellipsoid, is described by the eigenvalues $\lambda_i^2$ of the matrix $C_{ij}$, having the physical meaning of the dimensions of the cloud squared. 
This is in line with the analytic affine approach using ellipsoidal shape as an assumption \citep{LC86}.
The eigenvalues are found together with eigenvectors, whose orientation in both  coordinate systems (stellar and BH) changes with time.
The debris cloud is approximately symmetric with respect to the orbital plane, implying one of its eigenvectors is nearly vertical (orthogonal to the equatorial plane).
The order of the eigenvectors is as follows: the one closest to the $z$ axis (orthogonal to the plane of the initial orbit) is considered the third (and the corresponding eigenvalue is denoted as $\lambda_3$); from the other two, the eigenvector making the smallest angle with the propagation direction of the stellar core on the initial orbit (calculated as the tangent unit vector to the orbit, Eq.~\ref{E:parabolic:tangent}) is considered the first (and the corresponding eigenvalue denoted as $\lambda_1$), and the remaining one the second ($\lambda_2$). 

\begin{figure*}
\adjincludegraphics[width=1.0\textwidth,trim={1cm 1cm 0cm 0cm},clip]{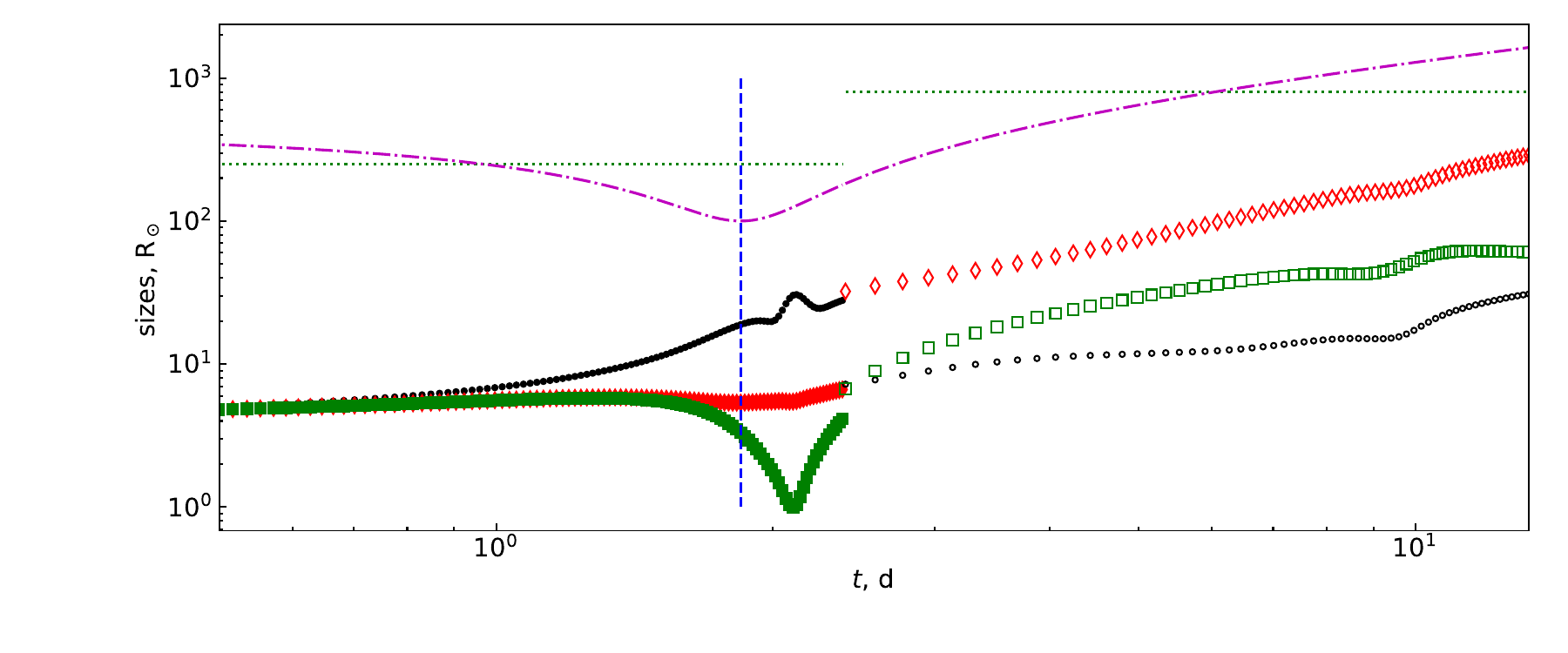}
 \caption{
 Evolution of the three principal spatial dimensions of the debris cloud $\lambda_i$ with time during the first two simulation stages. Solid symbols belong to \simD, open to \simDt. Green squares are $\lambda_3$ (vertical direction), black circles and red diamonds correspond, respectively, to  $\lambda_1$ (close to the propagation direction) and   $\lambda_2$. The sudden switch between $\lambda_1$ and $\lambda_2$ near $t=5$d is related to the gradual rotation of the principal axes with respect to the direction of motion. Horizontal dotted green lines are half-sizes of the simulation boxes, pericenter passage is shown with a vertical blue line. Purple dot-dashed line shows the distance between the star and the BH $r_*(t)$ as a function of time along the initial orbit.
 }\label{fig:shapes}
\end{figure*}

\begin{figure}
\adjincludegraphics[width=1.0\columnwidth,trim={0cm 0cm 1.5cm 0.5cm},clip]{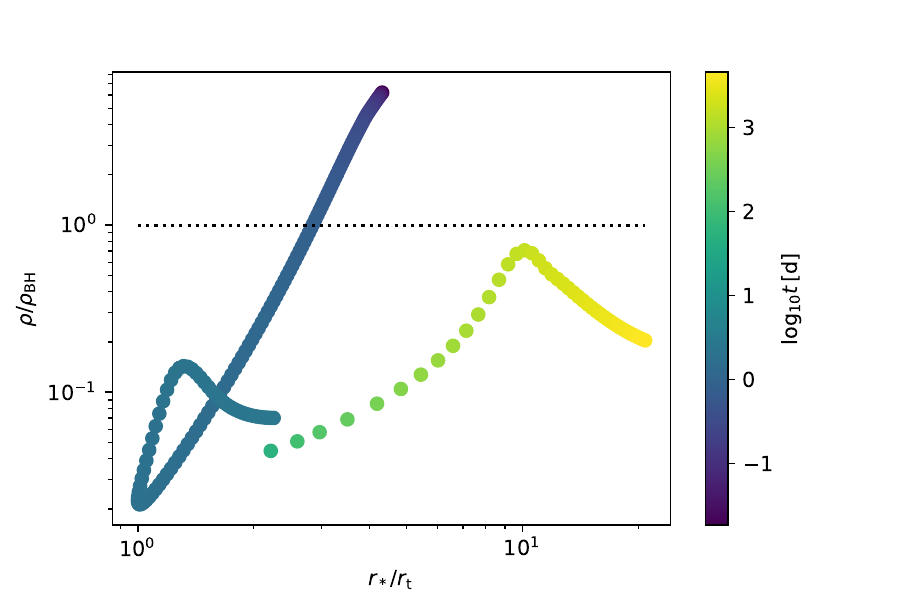}
 \caption{
Mean density (normalized by $\rho_{\rm BH} = M_{\rm BH} r_*^{-3}$) within the stellar debris (tracer condition $s> 0.5$) as a function of radial coordinate $r_*$ of the stellar center of mass throughout the simulation stages \simD\ and \simDt. 
 }\label{fig:rhomeanall}
\end{figure}

Fig.~\ref{fig:shapes} shows how these three eigenvalues change with time. 
The star is first (before the pericenter) stretched predominantly along the orbit.  
Then, apparently because of the pressure drop (maximal pressure within the star decreases by about an order of magnitude), the vertical dimension of the star collapses and experiences a bounce several hours after the pericenter passage. 
Further expansion makes the debris cloud highly anisotropic: expansion along the vertical and one of the horizontal directions is approximately linear, while in the remaining principal direction the size of the cloud remains $\sim 10\Rsun$. 
The smallest eigenvalue during the fly-by is $\lambda_2$, but after $t\sim 5$d the shortest dimension is the one closest to the tangent to the orbit.
The expansion anisotropy is apparently affected by self-gravity, making our simulation an example of a `self-gravitating pancake' \citep{2016MNRAS.455.3612C}.
The mean density of the stellar debris remains comparable but smaller than $M_{\rm BH} / R^3$, meaning that the BH tidal forces dominate over self-gravity in general, but the smallest dimension of the debris cloud is affected by self-gravity. 
The marginal importance of self-gravity in general is illustrated by Fig.~\ref{fig:rhomeanall} which compares the average density of stellar debris and the average density inside the sphere with the radius of $r_*$ (a similar approach was used by \citealt{2015ApJ...808L..11C}). 
The ratio $\rho / \rho_{\rm BH}$ shows the importance of self-gravity compared to tidal forces: above the dotted line in Fig.~\ref{fig:rhomeanall}, the star is gravitationally bound. 

\begin{figure}
\adjincludegraphics[width=1.0\columnwidth,trim={2cm 1cm 0cm 2cm},clip]{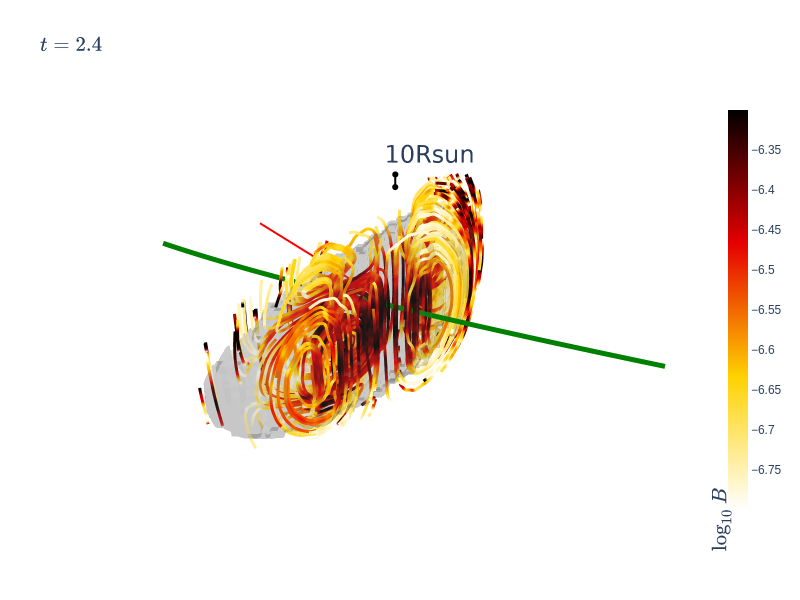}
 \caption{Three-dimensional rendering of the field lines at $t=2.4$d, stage \simDt. Color shows the field strength logarithm (code units), gray contours outline the region where $s=1/2$. Camera is set at an elevation of $60$deg, along the negative direction of the $Y$ axis. The orbit is shown with a solid green line, the red straight line is directed towards the black hole. The scale is shown by a 10\Rsun\ segment in the upper part of the picture.
 }\label{fig:flow2d}
\end{figure}

The magnetic field during this stage is being aligned with the flow. 
Magnetic flux conservation ensures that at large distances, the dominating magnetic field component is the one aligned with the second principal direction, along the debris cloud is extended the most. 
Just before the fallback begins, the magnetic field effectively consists of two loops in the vertical plane of the `pancake' (see Fig.~\ref{fig:flow2d}).

\begin{figure*}
\includegraphics[width=1.0\textwidth]{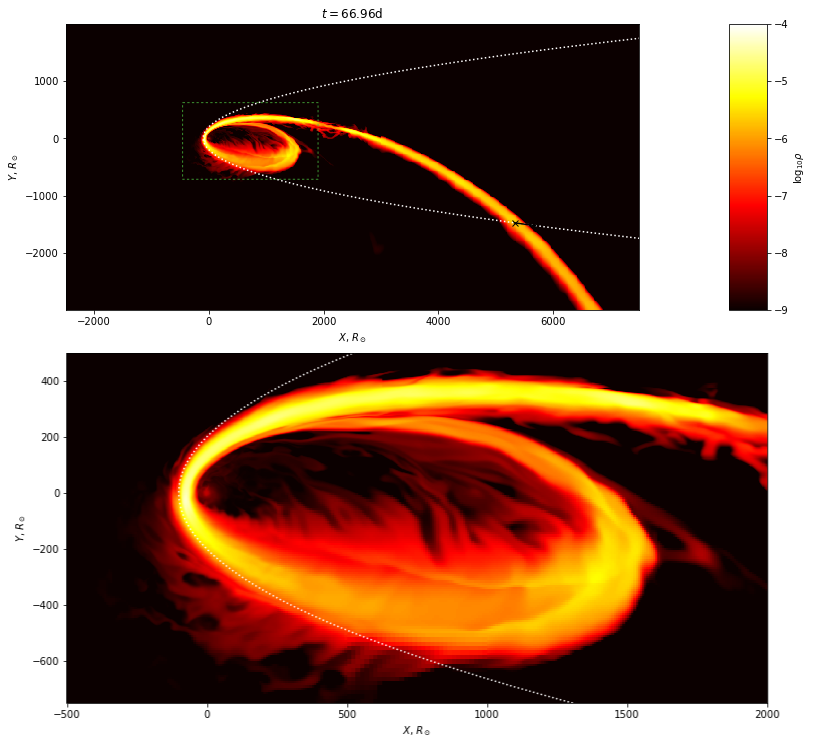}
 \caption{Equatorial-plane density map for $t\simeq 67$d, simulation \simHL. The upper panel shows the whole frame, and the lower panel is a zoom-in into the region around the black hole (shown with a dotted green rectangle in the upper panel). The initial orbit is shown with a dotted white line. For the development of the density distribution with time, see \url{https://youtu.be/DmWdy8LakR8}.}\label{fig:den}
\end{figure*}

\subsection{Fallback stream and nozzle shocks}

\begin{figure*}
\includegraphics[width=1.0\textwidth]{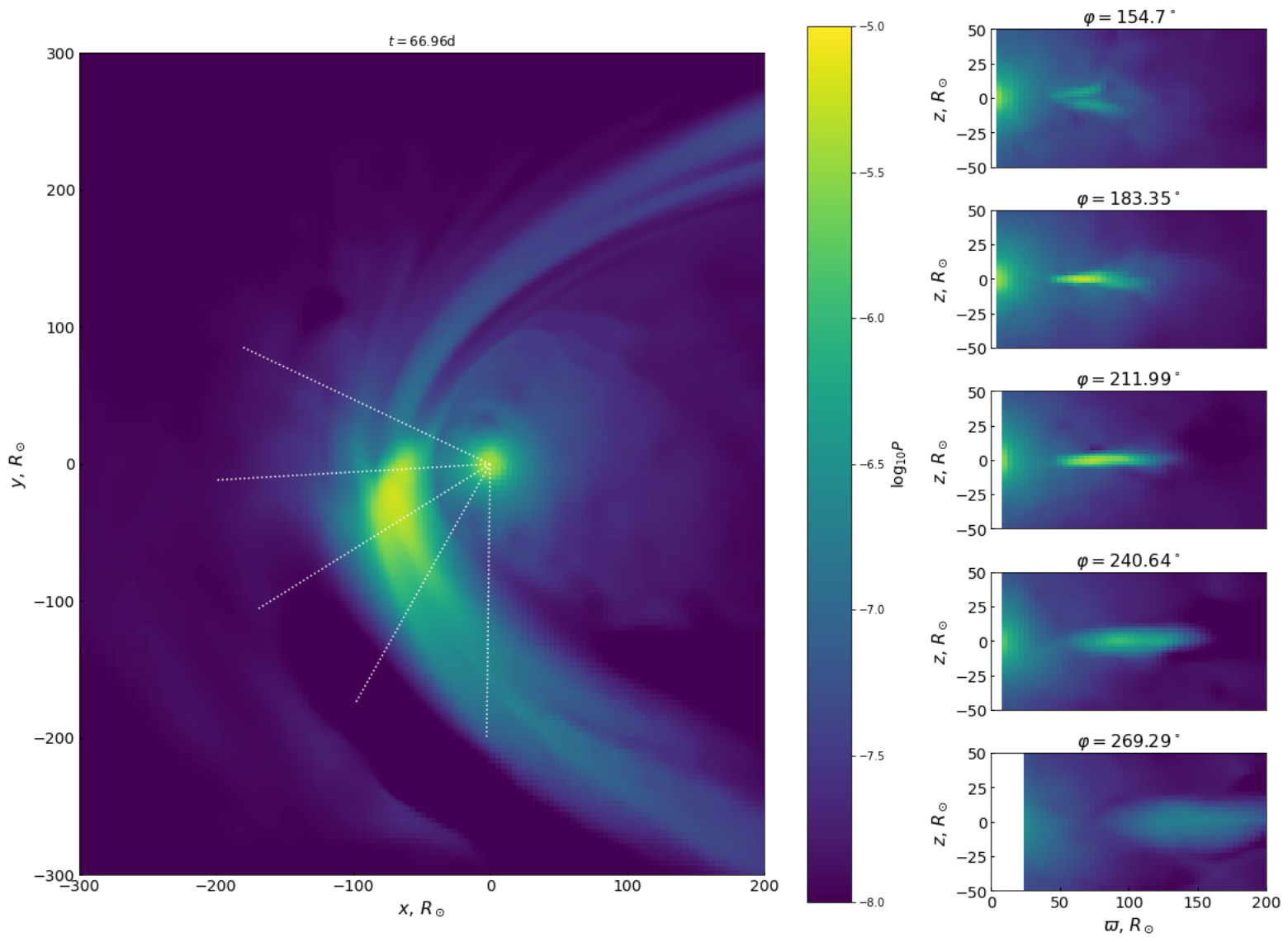}
 \caption{Pressure distribution in the equatorial plane (left panel) and in several meridional planes (right panels, the corresponding cross-sections are shown in the left panel with dotted white lines) covering the region of the pericenter and the nozzle shock. The color scale is logarithmic and identical in all the cross-sections. The simulation stage is \simHL, $t\simeq 67$d.}\label{fig:azcross}
\end{figure*}

The second passage starts with the fallback stream reaching the pericenter at $t\sim 30$d. 
During the following months, the flow is evolving into an eccentric ($e\sim 0.8$) vertically-extended ($H/r \sim 0.5$) structure aligned with the fallback stream. 
The relatively low eccentricity is a result of the low mass ratio of the simulation as $1-e \approx (M_*/M_{BH})^{1/3}$.
In Fig.~\ref{fig:den}, we show a density snapshot about a month after the beginning of the fallback.
The returning flow itself is moderately geometrically thick (its vertical scale $H \sim 0.1r$) and converges supersonically in the vertical direction, resulting in a system of shock waves near the pericenter of the initial orbit, which is referred to as nozzle shocks.
In Fig.~\ref{fig:azcross}, we show an equatorial cross-section of the pressure distribution in the flow near the pericenter and several meridional cross-sections that resolve a system of two oblique shock fronts visible as two pressure maxima above and below the equatorial plane in the first meridional cross-section. 
Further downstream, the two fronts merge into a narrow (in the vertical direction) overpressured region that further expands both vertically and horizontally.

\begin{figure}
\adjincludegraphics[width=1.0\columnwidth,trim={2cm 0cm 0cm 0cm},clip]{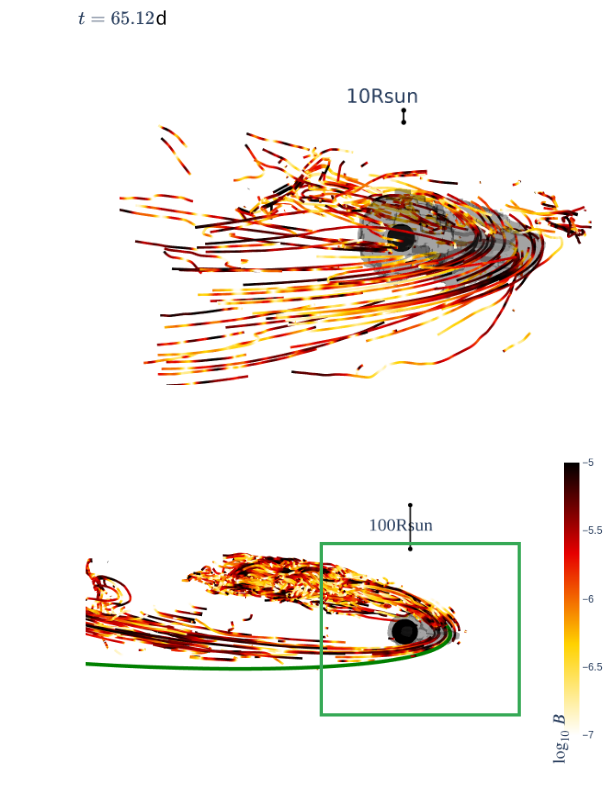}
 \caption{Pressure distribution (shown by transparent gray surfaces) and magnetic field lines plotted as 3D renderings for \simHL, $t\simeq 65$d. 
 Lower panel covers the region from $x = -500\Rsun$ to $x=1000$ in the BH-centered system of coordinates. 
 Upper panel shows a zoom-in into the central region, shown with a green box in the lower panel.
 }\label{fig:render65}
\end{figure}

Heating by the nozzle shocks is crucial for the subsequent evolution of the flow. 
First, it makes the flow geometrically thick (vertical scale $H \sim 0.5r$), that ultimately leads to the formation of a thick disk. 
Heating also creates transverse pressure gradients that broaden the distribution in angular momentum and in eccentricity.
Most importantly, the flow downstream of the nozzle shock becomes turbulent, which directly affects the shapes of the magnetic field lines.
Fig.~\ref{fig:render65} shows 3D renderings of the field line shapes. 
As one may see, the lines are much smoother upstream of the pericenter and become gradually more entangled downstream. 

\subsection{Circularization process}\label{sec:res:circ}

To track the process of circularization, we calculate the temporal evolution of the distribution of the stellar mass over the orbital eccentricity calculated for each cell as
\begin{equation}
       e = \sqrt{1 + \frac{2\varepsilon l^2}{G^2M_{\rm BH}^2}},
\end{equation}
where $\varepsilon$ is net mechanical energy, and $l$ is net angular momentum. 
For Keplerian motion, this expression gives the eccentricity of the bound or unbound free orbit. 
This is a Newtonian analogue of the expression used by \citet{andalman}. 

\begin{figure*}
\adjincludegraphics[width=1.0\textwidth,trim={0cm 0cm 3.5cm 0cm},clip]{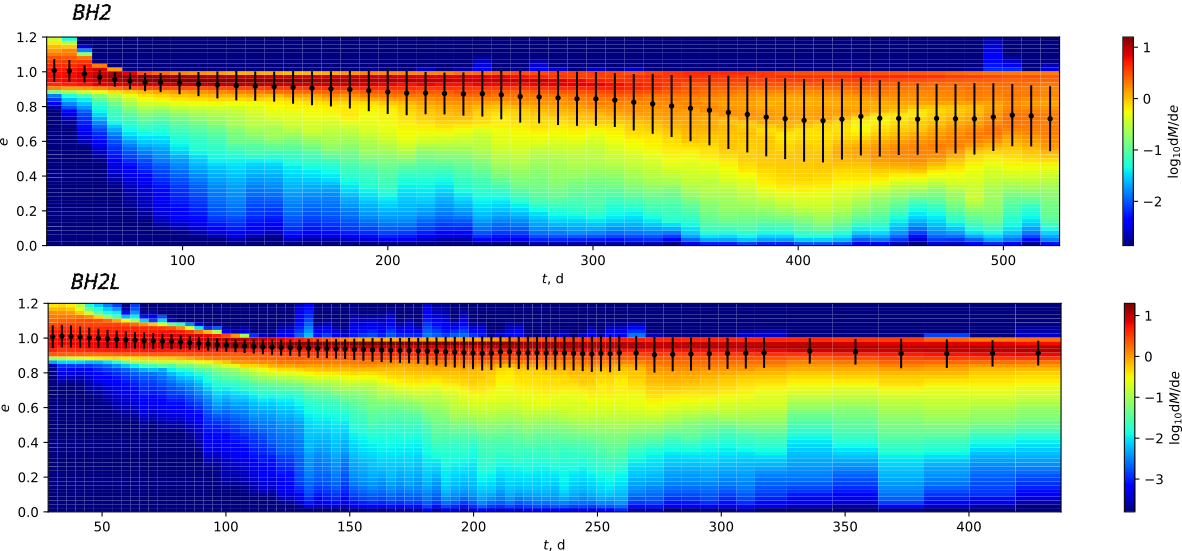}
 \caption{
    Eccentricity distribution in the runs \simH\ and \simHL\ after the formation of the fallback stream. Dots are average values, the error bars show root-mean-square deviations. 
 }\label{fig:edist}
\end{figure*}

Fig.~\ref{fig:edist} shows the distribution of the stellar mass $\int_ V s \rho \diff V$ over eccentricity in simulation \simH. 
Initially, the distribution is practically symmetric with respect to $e=1$, however by $t\sim 40-50$d the unbound material is mostly expelled from the box.
In the simulation \simH, the fallback stream in the eccentricity range $e\sim 0.9-1.0$ can be traced until $t\sim 300$d when it is quenched by the outer rim of the box (the fallback time for $a =5000\Rsun$ is about a year).
In the larger-box simulation \simHL, circularization runs slower, as there is an inflow of high-eccentricity material at later times, roughly until $t\sim 500$d. 
In both cases the matter passing through the nozzle shock forms a turbulent eccentric `accretion abalone' with $e\sim 0.8-0.9$. 
Internal shocks and collision with the fallback stream lead to further circularization.
While the resulting configuration (at $t\sim 500$d) is close to axial symmetry, the estimated eccentricity values $\sim 0.5-0.8$ are moderately high. 
This is expected for a thick, partially pressure-supported, turbulent toroidal structure, where motion is on average sub-Keplerian. 
We would hereafter refer to it as the disk.

\subsection{Magnetic energy and flux evolution}

To trace the evolution of the global properties of the magnetic field during all stages, we used the energies of the three magnetic field components in cylindrical geometry, calculated as tracer-weighed integrals over the simulation box
\begin{equation}\label{E:energy}
    E^{\rm M}_{\varpi,\, \varphi,\, z} = \frac{1}{2}\int B_{\varpi,\, \varphi,\, z}^2 s \diff V,
\end{equation}
where lower indices correspond to the coordinates of the cylindrical system centered on the BH (radial, azimuthal, and vertical, respectively).
Using the cylindrical system allows to separate the field components during the disk formation process where the conditions for toroidal, radial, and vertical fields are different. 

\begin{figure*}
\adjincludegraphics[width=1.0\textwidth,trim={2.5cm 0cm 3.5cm 0cm},clip]{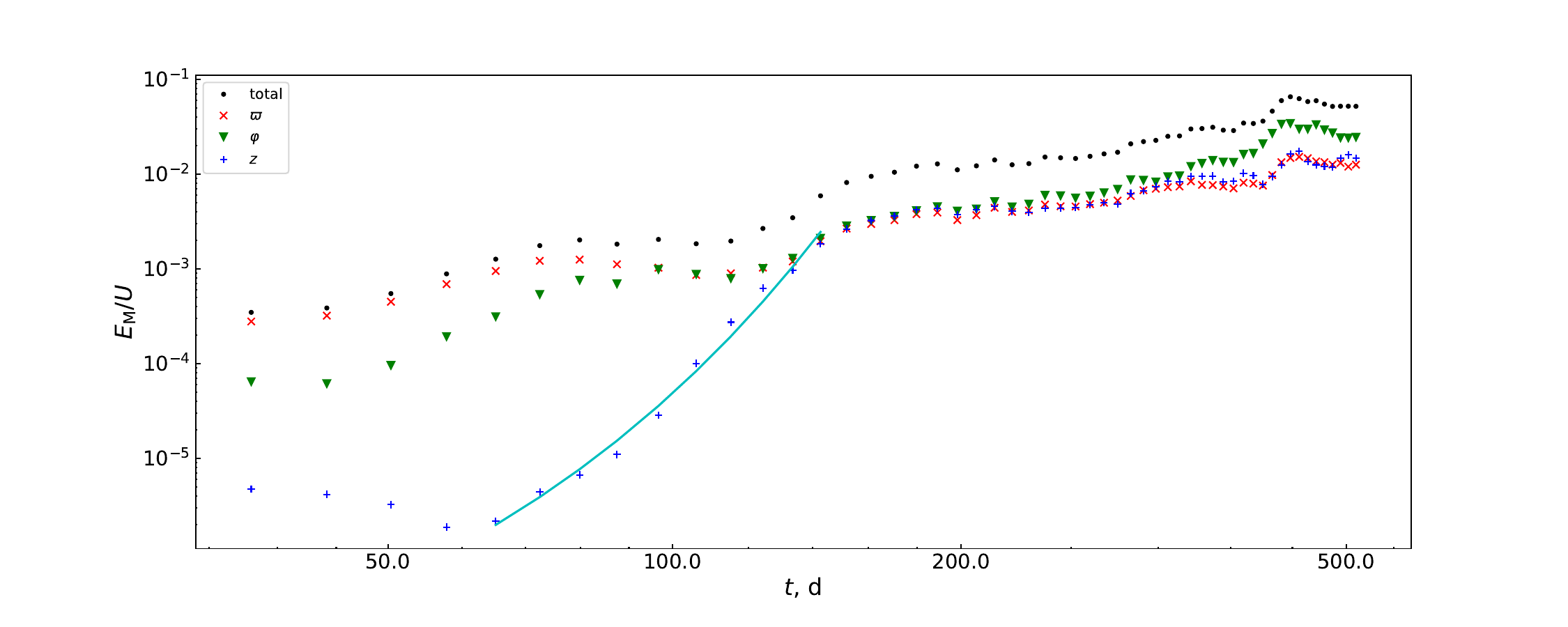}
\caption{
Magnetic field energies in the units of total heat (integrated over stellar material) as a function of time for the simulation \simH. Black dots are total $E^{\rm M}$, red x-s are $E^{\rm M}_\varpi$, green triangles are $E^{\rm M}_\varphi$, blue crosses are $E^{\rm M}_z$. The solid cyan line is an exponential fit to  $E^{\rm M}_z$ in the range $60{\rm d}<t < 150{\rm d}$.
 }\label{fig:bene}
\end{figure*}

In Fig.~\ref{fig:bene}, we show the evolution of energy in all the three magnetic field components during stage \simH. At the beginning of this stage, the magnetized fallback stream reaches the pericenter region and forms a nozzle shock at $t\sim 50$d. The decrease in $E^{\rm M}_z$ at $t \lesssim 60{\rm d}$ is likely related to a decrease in the vertical velocity component after the shock, leading to a decrease in the vertical magnetic field. After $t\sim 60$d, there is an increase in all the field components, most notably in $B_z$. The growth in the vertical field is well fitted by an exponent with an e-folding time of $11.0\pm 0.4$d (the fit is shown in Fig.~\ref{fig:bene}). 
Part of the overall magnetic energy growth trend is related to the loss of thermal energy, that is approximately constant before $t\sim 200$d and then decays by approximately an order of magnitude between $200$ and $500$d. 

\begin{figure*}
\includegraphics[width=1.0\textwidth]{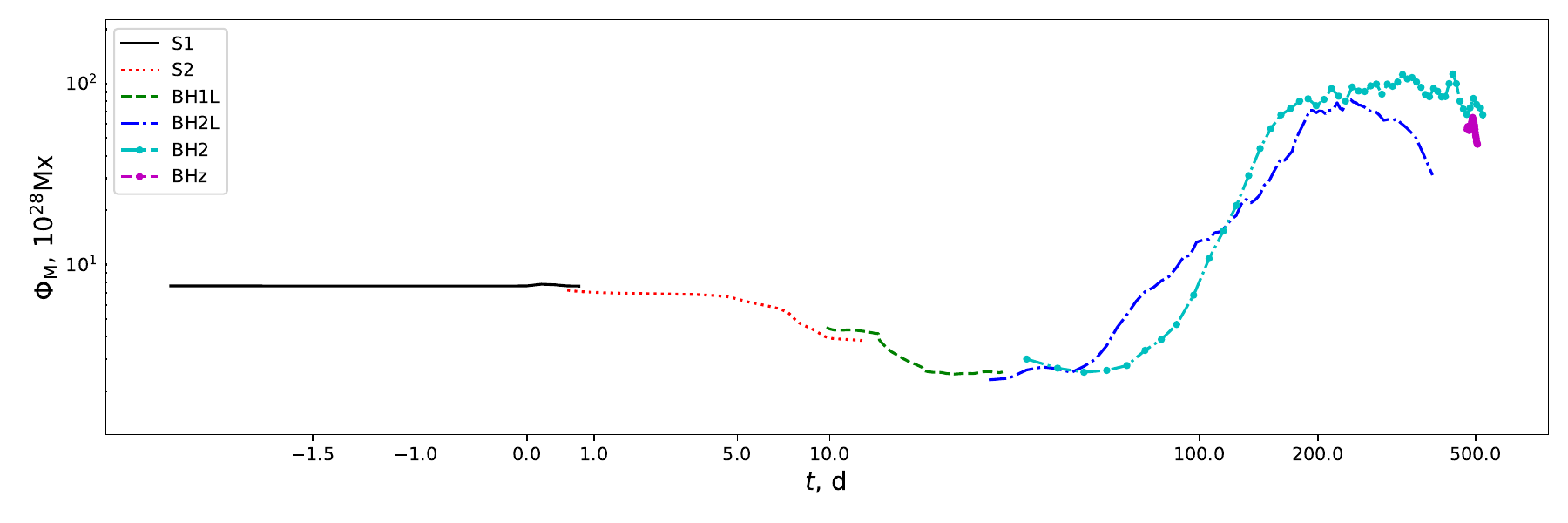}
 \caption{One-sided magnetic field flux as a function of time. Different curves represent different simulation stages: \simD\ (black solid), \simDt\ (red dotted), \simo\ (green dashed), \simHL\ (blue dot-dashed), \simH\ (teal dashed with thick dots) and \simHZ\ (thick solid magenta line).
 }\label{fig:phiz}
\end{figure*}

As a measure for magnetic field amplification, one may also use the magnetic flux that, unlike magnetic field energy, is a fundamentally conserved quantity. 
As far as one neglects magnetic field reconnection, this is also true for a one-sided flux (magnetic flux of a single polarity) defined as
\begin{equation}
    \Phi = \int \max\left( \vector{B} \cdot \vector{n}, 0\right) \diff A, 
\end{equation}
where integration is performed over the surface area $A$, and $\vector{n}$ is a unit vector normal to the surface. 
In our case, the total magnetic flux through any plane within the simulation domain is $0$ most of the time (the field lines do not cross the simulation box boundaries, and if they do, the fluxes of opposite signs compensate each other), hence a one-sided flux is much more useful.

In particular, we are using the vertical flux calculated through the orbital plane 
\begin{equation}
    \Phi_z = \int_{z=0} \max\left( B_z, 0\right) \diff x \diff y, 
\end{equation}
which initially equals $\sim 10^{28}\Mx$. 
As the flow has approximate symmetry with respect to the plane, one might expect this flux to be conserved, unless magnetic field is reconnecting or amplified. 
The one-sided flux is conserved well through the disruption process, decreases due to reconnection during the late expansion stage when the aspect ratio of the debris cloud is extremely large ($t\sim 5-30$d), and then gets amplified during the fallback stage ($t\sim 50-200$d).
All stages are shown in Fig.~\ref{fig:phiz}. The rapid amplification phase coincides with the rapid growth in magnetic energy, but the growth rate is expectedly about two times smaller. 
The differences in the evolution of magnetic energy and magnetic flux are mostly related to the vertical structure of the flow. 
The late gradual growth in energy seen in Fig.~\ref{fig:bene} is affected by the loss of internal energy as well as to the growth of the thick disk observed at later stages and the vertical expansion of the magnetized regions. 
The difference between the fluxes in \simH\ and \simHL\ shows the influence of the box size and resolution effects. 
In both cases, there is a considerable roughly exponential growth during the fallback stage, but the growth time varies by about a factor of 2. 
In both cases, the energy grows up to a saturation level of several per cent of the internal energy of the stellar debris.

\begin{figure*}
\adjincludegraphics[width=1.0\textwidth,trim={1.5cm 3cm 1.5cm 2cm},clip]{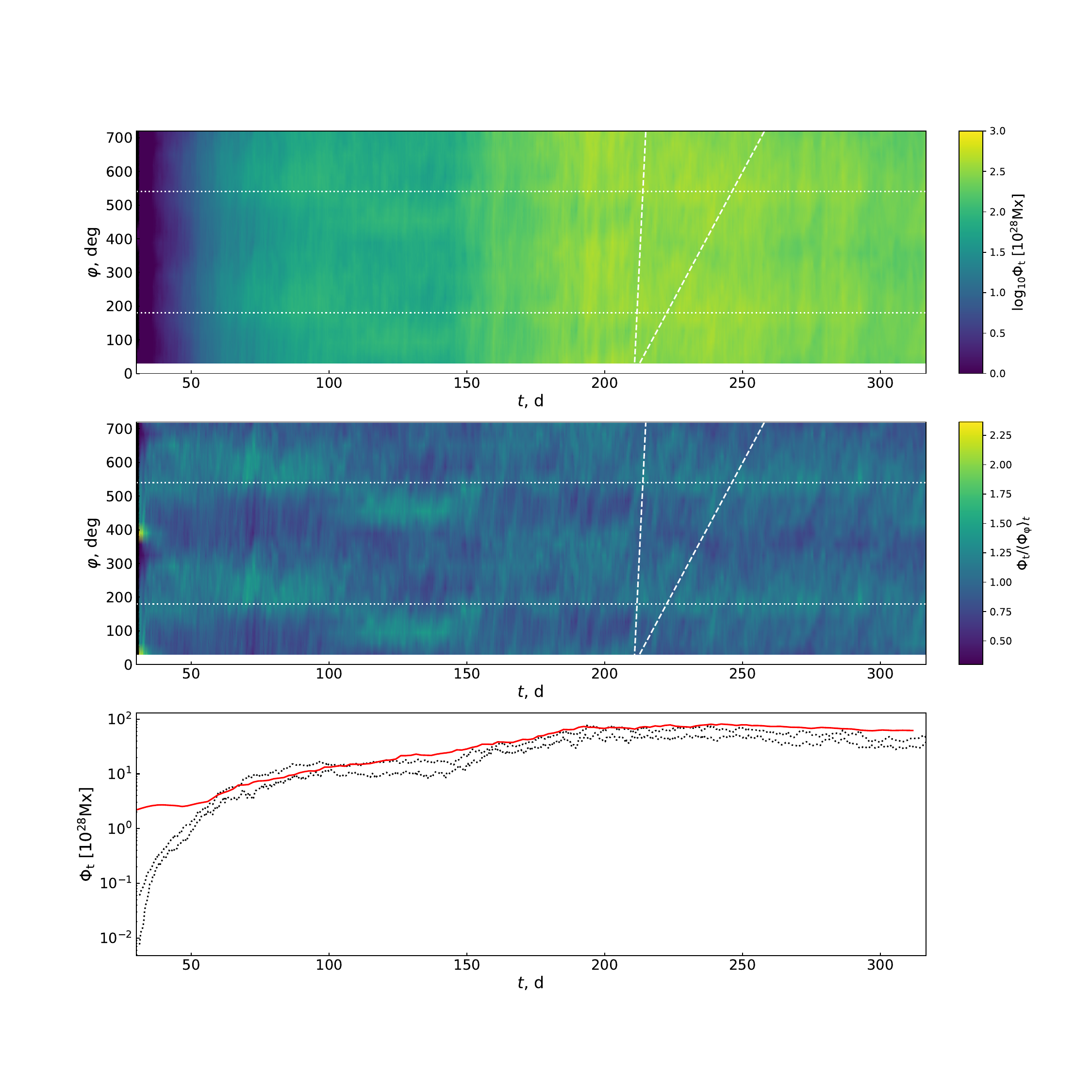} \caption{One-sided toroidal magnetic flux throughout \simH\ as a function of $\varphi$ and time. Upper panel shows the absolute flux values, middle panel shows the fluxes normalized by the azimuth-averaged value, and the lower shows the maximal and minimal $\Phi_{\rm t}$ (dotted black lines) compared to the vertical flux $\Phi_z$ shown with a solid red line. Dashed white lines in the upper two panels correspond to rotation at a rate $\Omega_{\rm K}(r_{\rm p}) = \sqrt{GM/r_{\rm p}^3}$ and $\Omega_{\rm K}(500\Rsun)$. White dotted lines highlight the azimuth of the pericenter. 
 }\label{fig:phit}
\end{figure*}

In a way similar to vertical flux through the equatorial plane, one can define a one-sided toroidal flux through a meridional plane at a particular azimuthal angle $\varphi$
\begin{equation}
    \Phi_{\rm t} =  \int_{\varphi = \rm const, \  \varpi < \varpi_{\rm max}} \max\left( B_\varphi, 0\right) \diff z \diff \varpi.
\end{equation}
In Fig.~\ref{fig:phit} we show the toroidal fluxes as functions of time for 11 such planes at different azimuthal angles. The maximal radius of integration is $\varpi_{\rm max} = 200\Rsun$. 
The flux depends much stronger on time than on $\varphi$ and behaves similarly to $\Phi_z$. 
For most of the time, the toroidal flux is larger downstream of the pericenter, suggesting amplification in the vicinity of the nozzle shock. After passing through the nozzle shock, the flow expands dramatically in vertical direction, which decreases the average value of the toroidal field but does not change the flux. 
At later times, it is possible to see Keplerian motions at the outer edges of the integration aperture as the field loops are advected along with the rotating flow. 
Both types of magnetic fluxes, vertical and toroidal, grow together, meaning that the dramatic increase of $E^{\rm M}_z$ during the fallback is not entirely caused by magnetic energy redistribution over components, as all the components of the field are growing simultaneously. 

\subsection{Magnetic fields in the disk}

\begin{figure}
\includegraphics[width=1.0\columnwidth]{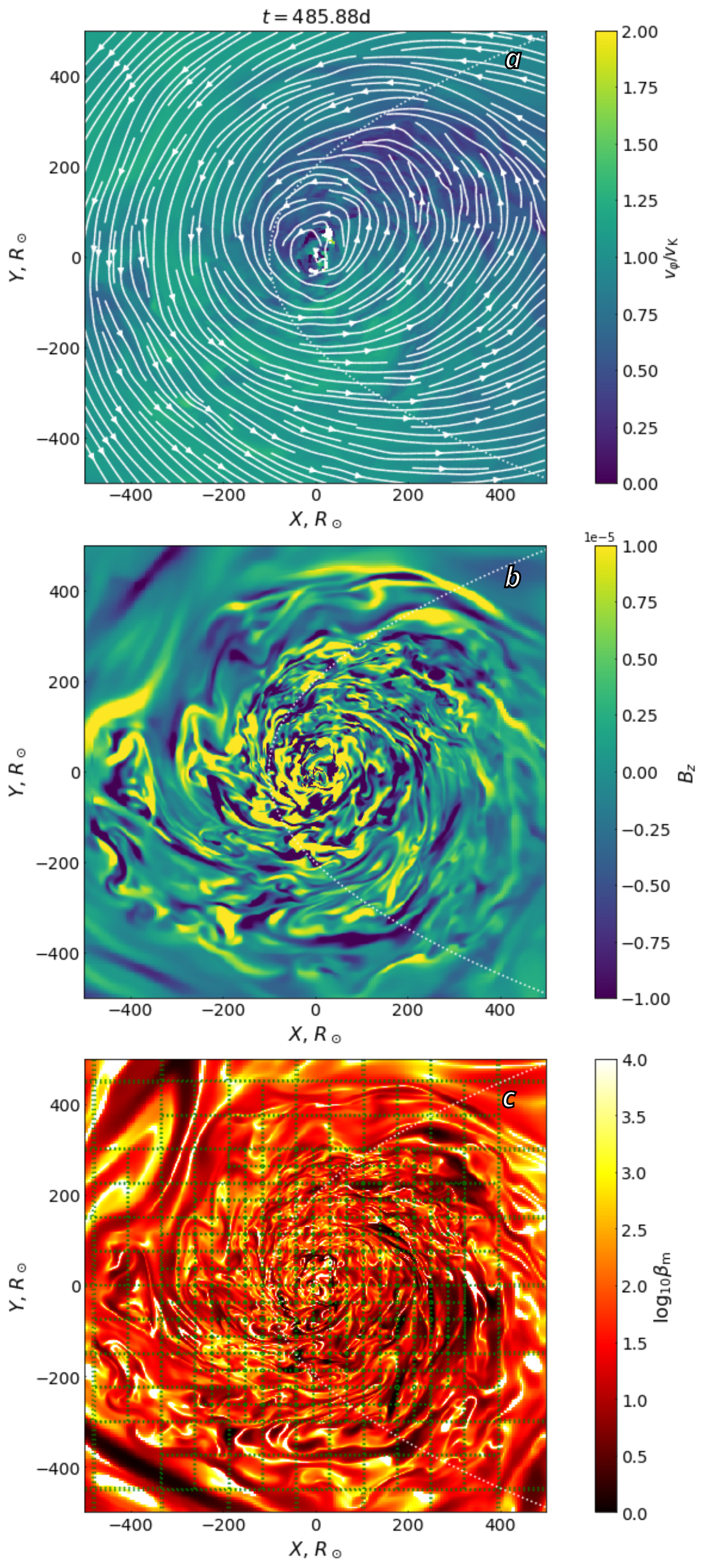}
 \caption{
 Equatorial cross-sections of rotation velocity in Keplerian units ($a$), vertical magnetic field ($b$), and $\beta_{\rm m}$ ($c$). In panel $a$, velocity stream lines are overplotted with solid white lines. Dotted green lines in panel $c$ are the boundaries of mesh blocks. Simulation \simHZ, $t \simeq 485$d. Dotted white lines show the initial orbit. 
 }\label{fig:cauliflower}       
\end{figure}

Independently of the simulation box size, the vertical magnetic flux reaches a maximal value of $\Phi_z \sim 10^{30}\Mx$ at $t\sim 200$d. This was seen in both \simH\ and \simHL\ 
runs. 
For the \simH, this is also the time when the fallback stream is quenched. 
To resolve the evolution of the field in the disk, we added stage \simHZ\ with one additional level of refinement. 
After the quenching of the fallback stream, magnetic field components remain in approximate equipartition with each other. 
Total magnetic energy is stabilized at a level about 4 per cent of the total internal energy. 
The overall change in $\beta_{\rm m}$ is thus about a factor of 20. 
In the time range $t = 470-500$d (model \simHZ) different components of the field contribute to the energy as
\begin{equation}
    E^{\rm M}_z = \left( 0.0136 \pm 0.0008\right) U,
\end{equation}
\begin{equation}
    E^{\rm M}_\varphi = \left( 0.031 \pm 0.002\right) U,
\end{equation}
\begin{equation}
    E^{\rm M}_\varpi = \left( 0.0139 \pm 0.0016\right) U,
\end{equation}
where $U$ is total internal energy. 
All the energy components were calculated using the tracer as a weighing factor (Eq.~\ref{E:energy}).
The result was averaged over 39 frames, and the uncertainties in the estimates above are the root-mean-square variations between the frames. 
Local magnetization differs considerably over the frame, reaching levels of $\beta_{\rm m} \sim 1$ in particularly strongly magnetized regions. 
In Fig.~\ref{fig:cauliflower}, we show the equatorial cross-sections of the disk revealing its complex chaotic structure and the variations of $\beta_{\rm m}$. 
Velocity map shows that rotation direction in the orbital plane is consistent, and the rotation velocity $v_\varphi$ is of the order local Keplerian velocity. 
The magnetic fields are entangled but have preferential directions and typical correlation length scales that may be probed with Fourier analysis (as will be done later in Section~\ref{sec:res:fou}).

\begin{figure}
\includegraphics[width=1.0\columnwidth]{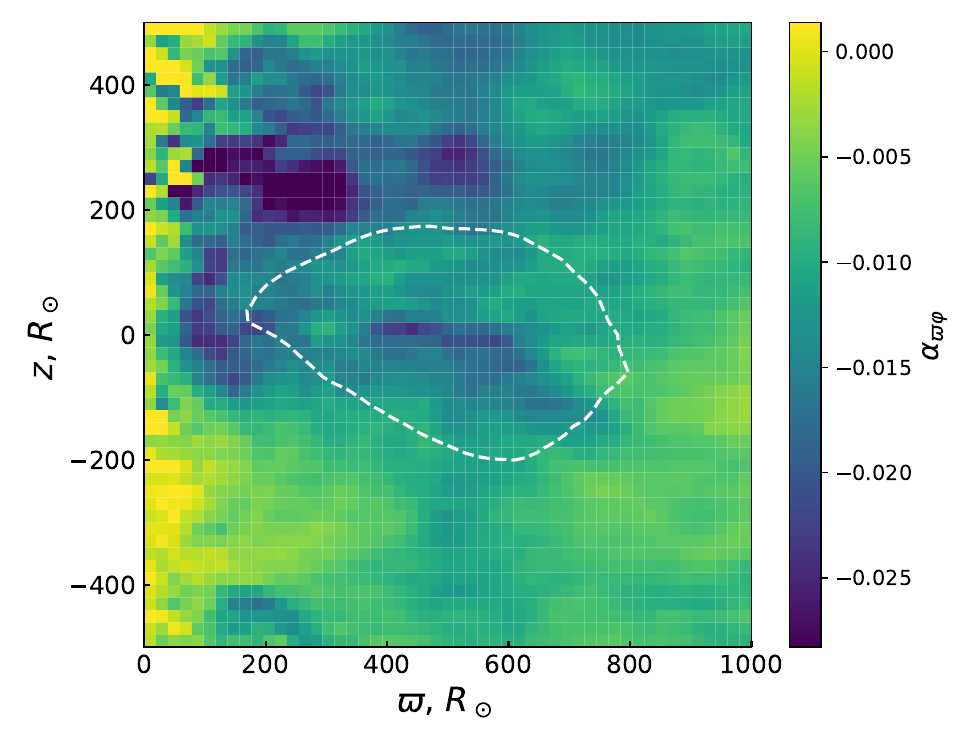}
\includegraphics[width=1.0\columnwidth]{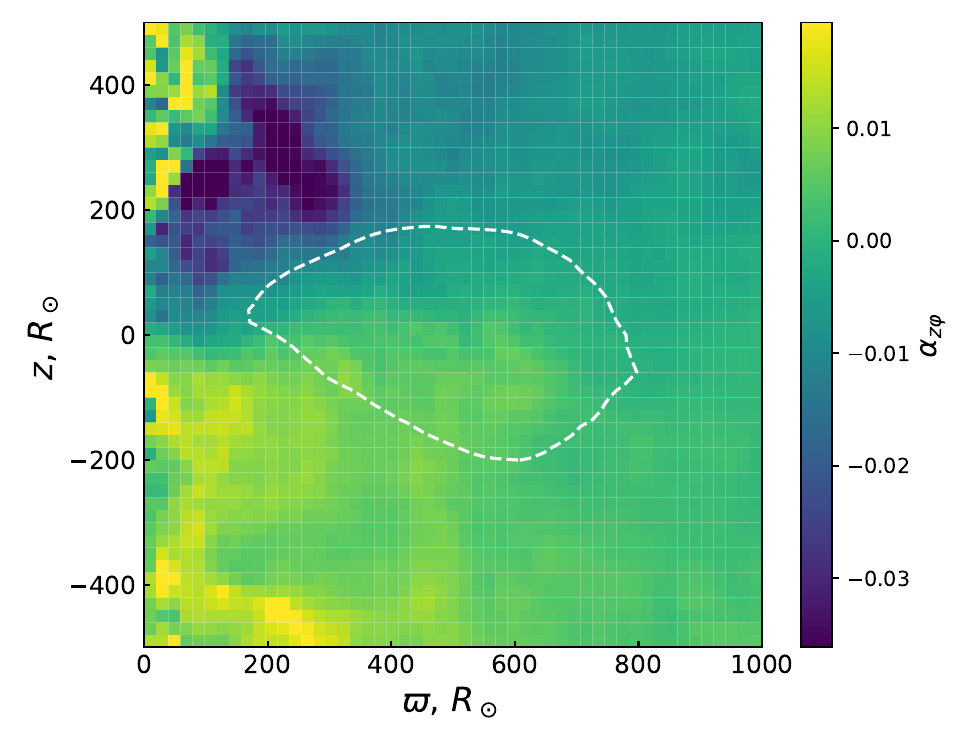}
 \caption{Magnetic field correlations over azimuth and time ($t = 480-520$d, 50 frames) normalized over the average local pressure. The dashed white contour is the isoline of the averaged density equal to one half of its maximal value. 
 }\label{fig:alphas}
\end{figure}

The magnetized disk in the stage \simHZ\ mostly has the highest level of AMR, meaning its thickness is resolved at a factor of $H/ \Delta z \gtrsim 100$. 
Thus, we expect that the linear effects in magnetic field evolution (such as turbulent dynamo) are well resolved. 
For the non-linear processes, this is not that obvious. 
An important criterion here is the quality factor \citep{2010ApJ...711..959N,2011ApJ...738...84H}
\begin{equation}
    Q = \frac{2\uppi v_{\rm A}}{\Omega_{\rm K} \Delta z},
\end{equation}
where $v_{\rm A} =  B / \sqrt{\rho}$  is Alfv\'{e}n velocity, $\Omega_{\rm K}$ is Keplerian frequency. 
This quantity is usually defined for different magnetic field components individually, but we will use the absolute value, as the fields are already entangled and nearly isotropic when they become dynamically important. 
At the highest level of resolution and for a tracer value $s > 0.5$, the typical values of the quality factor are $Q \sim 30-100$ near the equatorial plane, meaning that the non-linear processes affecting magnetic field evolution such as Magneto-Rotational Instability (MRI, see \citealt{MRI}) are well resolved. 
During the earlier stages, large $\beta_{\rm m}$ values make it difficult to ensure that the quality factor is large enough (typically, $Q\sim 20$ within the fallback stream), but one would expect the magnetic fields to be dynamically unimportant. 

Averaging over the azimuthal angle and several consecutive frames of the high-resolution simulation \simHZ\ allows one to calculate the mean field values, their dispersions and correlations. 
In Fig.~\ref{fig:alphas}, we show the normalized correlations of the cylindrical magnetic field components calculated as
\begin{equation}
    \alpha_{\varpi\varphi} = \frac{1}{\langle P \rangle} \left\langle (B_\varpi-\langle B_\varpi \rangle) (B_\varphi - \langle B_\varphi\rangle)\right\rangle 
\end{equation}
and 
\begin{equation}
    \alpha_{z\varphi} = \frac{1}{\langle P \rangle} \left\langle (B_z-\langle B_z \rangle) (B_\varphi - \langle B_\varphi\rangle)\right\rangle,
\end{equation}
where the angular brackets correspond to spatial (in a certain small range of cylindrical radii $\varpi$ and $z$ and within the entire available range of $\varphi$) and temporal (50 frames covering the time span $t= 480-520$d) averaging of a given quantity. 
Tracers are not involved at this stage.
The former correlation quantity $\alpha_{\varpi\varphi}$ coincides by definition with the dimensionless magnetic viscosity parameter introduced by \citet{1973A&A....24..337S} and responsible for radial transfer of angular momentum within the disk. 
Negative $\alpha_{\varpi\varphi}$ corresponds to outward transport of angular momentum. 
The other quantity, $\alpha_{z\varphi}$, describes the angular momentum transfer in vertical direction, and its sign (negative in the upper hemisphere, positive in the lower) shows that angular momentum is predominantly transported away from the equatorial plane by magnetized outflows \citep{1982MNRAS.199..883B}. 
In the figure, we also show an isoline of the density $\langle \rho \rangle$ averaged the same way. 
The shape of the contour gives an impression about the size, shape, and thickness of the disk.
The typical values of both $\alpha_{\varpi\varphi}$ and $\alpha_{z\varphi}$ in the inner parts of the disk are about $2-3$ per cent. 
Typical Reynolds stresses are one-two orders of magnitude larger, suggesting that magnetic fields do not play as important a role as turbulence in angular momentum redistribution. 
Dispersions in Reynolds stress components are large and do not allow to make any conclusive statement about the direction of angular momentum transfer by turbulence. 

\subsubsection{Power-density spectra}\label{sec:res:fou}

\begin{figure*}
\includegraphics[width=1.0\textwidth]{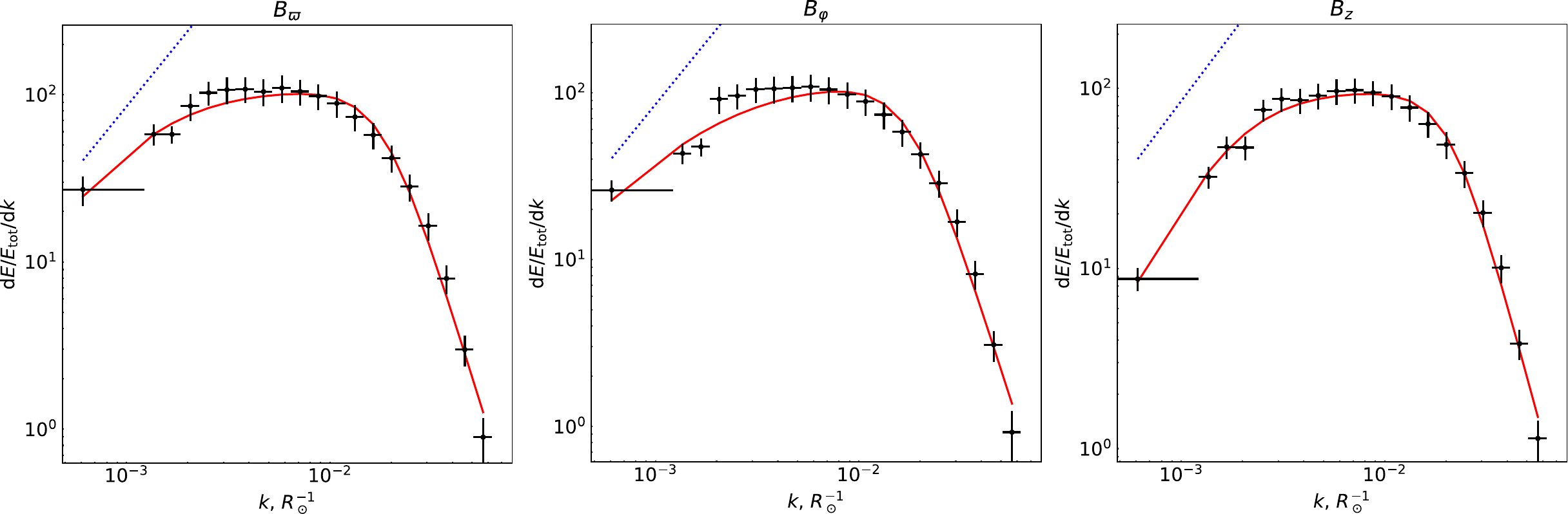} 
 \caption{PDSs of the three magnetic field components, $B_\varpi$, $B_\varphi$, and $B_z$ averaged over 50 frames of the \simHZ\ simulation. 
 The vertical error bars show the root-mean-square variation within the bin and between the frames. Solid red lines show the best fits with the empirical Eq.~(\ref{E:fnorm}).
 The dotted blue lines show Kazantsev's power-law scaling $\propto k^{3/2}$ with identical (arbitrary) normalization.}\label{fig:PDS}
\end{figure*}

To probe the scales and structure of the magnetic fields at the late stages, we have taken individual field components from 50 snapshots of the high-resolution simulation \simHZ\ (the same as used for the azimuthally averaged plots in the previous section), interpolated them on a regular mesh in a $10^3\times 10^3 \times 10^3\Rsun$ box centered on the black hole, and calculated three-dimensional Fourier spectra. 
The regular grid had $128$ elements in each direction that implies a spatial resolution $\sim 10$, two times finer than the zeroth AMR level, meaning that for the disk interior, the maximal resolution is comparable to or better than the Nyquist limit. 
The three-dimensional spectra then were averaged over time and directions in 20 logarithmically-spaced bins in wave number $k$. 
Wave number is here defined as linear spacial frequency, assuming $\e^{2\uppi \i k x}$ dependence on the coordinate for a harmonic wave moving in $x$ direction. 
In Fig.~\ref{fig:PDS}, we show the average isotropic power-density spectra (fraction of total power per unit wave number $k$). 
The uncertainties were calculated as the root-mean-squared errors within the wavenumber bin divided by the square root of the number of points and include variations both in time and in wave vector space.

The spectra are nearly flat in the range of wave numbers $k \sim 0.003-0.01$ and steeply decrease at higher $k$. 
The rise of the spectrum at low wave numbers has a slope of $\propto k^{1-2}$.
Fitting the spectral shape for $B_z$ with a function
\begin{equation}\label{E:fnorm}
    f = \frac{f_{\rm norm}}{1 + \left(k/k_1\right)^{-p} + \left(k/k_2\right)^{q}},
\end{equation}
where $p>0$ describes the low-$k$ behavior, and $q>0$ is responsible for the high-$k$ decrease in the PDS, yields a reasonable fit for all the components (see Table~\ref{tab:pdsfit}). 
For all the components, the lower break $k_1$ occurs at about the size of the disk ($1/k_1 \sim 500-600\Rsun$). 
The second break is at about half of the the pericenter distance and is probably related to the size of the fallback stream or the nozzle shock. 
The strongest differences in the PDS are at the lowest frequencies, where $B_\varphi$ and $B_\varpi$ clearly dominate over the vertical field. 
As a result, the best-fitted slopes $p$ are considerably smaller for $B_\varphi$ and $B_\varpi$. 

\begin{table*}\centering
\caption{Cylindrical magnetic field component fitting with the empirical model (Eq.~\ref{E:fnorm}).}\label{tab:pdsfit}
\begin{tabular}{lcccccc}
\hline \hline
 & $f_{\rm norm}$ & $k_1$, $\Rsun^{-1}$ & $k_2$, $\Rsun^{-1}$ & $p$ & $q$ & $\chi^2$\\
$B_z$ & $100\pm 6$ & $(1.70\pm 0.15)\times 10^{-3}$ & $(2.10\pm 0.10)\times 10^{-2}$ & $2.12\pm 0.18$ & $4.2\pm 0.2$ & 0.8\\
$B_\varphi$ & $120\pm 30$ & $(1.8\pm 0.8)\times 10^{-3}$ & $(1.7\pm 0.2)\times 10^{-2}$ & $1.2\pm 0.4$ & $3.8\pm 0.3$ & 1.2 \\
$B_\varpi$ & $113\pm 17$ & $(1.1\pm 0.3)\times 10^{-3}$ & $(1.75\pm 0.16)\times 10^{-2}$ & $1.5\pm 0.5$ & $3.9\pm 0.3$ & 0.6 \\
\hline 
\end{tabular}
\end{table*}

The measured PDS shapes are interesting to compare to the spectra of MHD turbulence produced by turbulent dynamo in numerical simulations \citep{2012SSRv..169..123B}.
The early stages of magnetic field amplification are characterized by a broadband noise with a peak at several inverse injection scales and a low-wavenumber scaling following the Kazantsev law $\propto k^{3/2}$ \citep{1968JETP...26.1031K}. 
Kazantsev scaling is consistent with the low-$k$ slope estimated for $B_\varphi$ and $B_\varpi$ and marginally consistent with the $p\sim 2$ slope found for $B_z$. 
With time, the spectrum is expected to become broader and develop an inertial interval with the Kolmogorov $\propto k^{-5/3}$ scaling, at least in the case of zero net helicity (see figure 1 in \citealt{2012SSRv..169..123B}). 
The saturation value for the magnetic field energy density is different than those in MHD turbulence simulations \citep{2009ApJ...693.1449C} that is probably related to the differences in the physical setup: boundary conditions, inhomogeneity and anisotropy of the flow as a whole. 
The magnetic field amplification time scales in these simulations are about an order of magnitude longer than the dynamical time scale, broadly consistent with the time scale seen in our simulation. 

\section{Discussion}\label{sec:disc}

\subsection{Initial magnetic flux}

We have assumed a simple dipole-like magnetic field confined to the interior of the star. 
Such a field is impossible to observe directly, though it should affect the pulsational properties of the star \citep{2015Sci...350..423F}. 
The strongest fields reported for stellar interiors, to our knowledge, are the $\sim 10^4-10^6$G fields in the cores of giant stars measured by \citet{2023A&A...670L..16D} using internal gravity modes. 
Main-sequence stars normally have surface large-scale fields in the range $1-10^4$G (see \citealt{2022A&A...662A..41R} for a review on magnetized low-mass stars), that implies vertical one-sided magnetic fluxes $\Phi_* \sim 10^{22}-10^{26}$Mx, between $10^{-6}$ and $10^{-2}$ of the flux in the model considered here. 
For the Sun, in-situ measurements estimate the total open photospheric magnetic flux of $\sim 10^{24}$Mx \citep{2008JGRA..11312103O}. 
If the exponential growth stage observed at $t\sim 100$d is ubiquitous, it would require $100-200$d to reach equipartition between the magnetic field and internal energy densities, starting with a realistic initial magnetization. 
This time scale is comparable to the circularization time scale observed in our simulation, on one hand, and to the life time of most TDEs, on the other.
Hence, magnetic field has enough time to be amplified to the values that might affect the dynamics of the disk. 

Most stars, however weak are the magnetic fields on the surface, have magnetized winds \citep{2015A&A...577A..28J}, potentially rotating, where tangential field components decrease with distance $\propto 1/R$ (Parker scaling, see \citealt{1958ApJ...128..664P}). 
The flux contained in this field exceeds the flux within the star by a linear factor 
\begin{equation}
    \Phi_{\rm w} \simeq 2\uppi B_{\rm A} \int_{R_{\rm A}}^{R_{\rm max}} \frac{R_{\rm A}}{R} R \diff R \sim 2\uppi B_{\rm A} R_{\rm A}R_{\rm max},
\end{equation}
where $R_{\rm A}$ is the radius of the Alfv\'{e}n surface of the magnetized wind, and $B_{\rm A}$ is the field strength at this radius. 
This estimate is valid for both tangential components (polar and azimuthal) under the assumption of constant wind speed, whereas rotation additionally enhances the toroidal component. 
Turbulent motions in the wind and its variability convert some part of the tangential magnetic field into radial \citep{2008JGRA..11312103O,2013LRSP...10....5O}. 
For the Solar wind, the magnetic field strength at $R \sim 1$AU is $\sim 5\times 10^{-5}$G, implying a magnetic flux of $\Phi_{\rm w} \sim 10^{23} \left(R_{\rm max}/{\rm AU}\right) \Mx$. 
The outer radius is determined either by dissipation in the wind or by the history of wind ejection and likely exceeds the size of the Solar system by a considerable factor. 
As a result, a typical main-sequence stellar wind contains more magnetic flux than the star itself.
Similar reasoning is applicable to the magnetic energy carried by the wind. 

The velocity of a solar-type star wind ($300-600\kms$) is normally much smaller than the relative velocity during the fly-by
\begin{equation}
\begin{array}{l}
\displaystyle    v_{\rm fb} = \sqrt{\frac{2GM_{\rm BH}}{R_{\rm t}}} = \sqrt{\frac{2GM_{\rm BH}^{2/3} M_*^{1/3}}{R_*}} \\
\displaystyle \sim 6\times 10^4 \left( \frac{M_{\rm BH}}{10^6 \Msun}\right)^{1/3}  \left( \frac{M_*}{\Msun}\right)^{-1/6}  \left( \frac{R_*}{\Rsun}\right)^{-1/2} \kms,\\
\end{array}
\end{equation}
meaning that the wind is tidally disrupted in a similar manner to the star itself.
At least some part of the wind gets captured by the BH gravity earlier and more efficiently than the bulk of the stellar material.
The likely effect of the early accretion from the magnetized stellar wind is formation of a highly magnetized low-accretion-rate disk that can work as a reservoir for the magnetic flux during the fallback and facilitate the circularization process via the `runaway circularization' scenario proposed by \citet{andalman}. 
We plan to include the wind fields in subsequent versions of our model. 

\subsection{Magnetic flux and jet formation}

As we have shown, the stellar magnetic field is amplified to an equipartition state where magnetic energy is several per cent of the internal energy. 
The fraction of this energy that may be used to power a relativistic jet depends on how efficiently the fields are transported towards the inner, relativistic regions of the disk.
The difference in spatial scales between the simulated disk size ($\gtrsim r_{\rm p} = 100\Rsun$) and the last stable orbit\footnote{for the BH mass $M_{\rm BH} = 3\times 10^3\Msun$, $R_{\rm ISCO} = (1-9)GM_{\rm BH}/c^2 \sim (0.01-0.1)\Rsun$, depending on the Kerr parameter of the BH} is more than three orders of magnitude. 
The small-scale loops existing at $t\sim 500$d are going to open up because of differential rotation, creating patches of magnetic fields piercing the disk and extending to infinity. 
When the open field lines pierce the vicinity of the horizon, one would expect jet formation via Blandford-Znajek mechanism \citep{BZ}, reaching the power of about $\dot{M}c^2$ \citep{2011MNRAS.418L..79T} for the case of a rapidly rotating BH. 

For Eddington-limited mass accretion rate, the jet power is basically determined by the BH mass. 
On the other hand, it depends on the magnetic flux through the BH horizon, that may be estimated for the MAD (magnetically arrested disk) limit using approximate balance between the pressure in the disk and magnetic pressure inside the last stable orbit. 
This leads to an estimate for the magnetic flux (marking the transition towards magnetically arrested disk for an about-Eddington accretion flow, see \citealt{2014ARA&A..52..529Y, 2016MNRAS.462..960S})
\begin{equation}
\begin{array}{l}
  \displaystyle  \Phi_{\rm Edd} \simeq \uppi \left( \frac{GM_{\rm BH}}{c^2}\right)^2 B_{\rm ISCO}
    \sim \frac{GM_{\rm BH}}{c^2} \sqrt{\frac{GM_{\rm BH}}{\varkappa}} \\
   \displaystyle \qquad{} \sim 10^{24} \left( \frac{M_{\rm BH}}{3\times 10^3\Msun} \right)^{3/2} \Mx, \\ 
\end{array}
\end{equation}
meaning that the initial magnetic flux contained within a solar-type star is comparable to the flux required for efficient jet launching for the BH mass adopted in our simulations. 
A massive BH requires larger $\Phi_z$ and thus a longer episode of magnetic field amplification. 
In the presence of rapid exponential field amplification stage, this obstacle is easy to overcome. 
A potentially more serious issue might be the structure and topology of the fields reaching the last stable orbit. 
To understand the further evolution of magnetic fields, more simulations are needed focusing on the inner, relativistic parts of the accretion disk.

\section{Conclusions}\label{sec:conc}

We report the first ever MHD simulation of a TDE from an initially magnetized star towards the formation of an accretion disk. 
The initial dipole-like field of the star first gets deformed and aligned with the maximal stretch direction. 
Then, a magnetized fallback stream returns to the vicinity of the BH, passes a system of oblique nozzle shocks, and forms an eccentric geometrically thick turbulent accretion flow, where the magnetic fields are amplified exponentially on a time scale of approximately the local Keplerian time. 

The magnetic field energy growth stops at about several per cent of the internal energy of the stellar debris, suggesting that amplification saturates at some sort of equipartition. 
The average value of magnetic beta within the disk is of the order several tens. 
The amplification of the magnetic fields close to the BH is compensated by the losses of magnetized matter in weakly collimated outflows. 

The magnetic fields established in the thick accretion disk during the circularization process are chaotic and show a preferential correlation length somewhat smaller than the pericenter radius. 
Their power density spectra are consistent with the early stages of turbulent dynamo amplification. 

In the following papers, we are planning to further explore the parameter space, considering stars with realistic structure and equation of state (radiation pressure is probably a particularly important factor), increasing the mass ratio, and varying the properties of the initial magnetic field. 
Apart from all the conventional physics, we are planning to include the magnetized wind of the star that we expect to be an essential constituent of the early evolution of a TDE.

\begin{acknowledgments}
    AL thanks James Beattie for enlightening discussions,  and the Canadian Institute for Theoretical Astrophysics for their warm hospitality and support.
    PA thanks Yannis Liodakis and Adelle Goodwin for the insights about the observational properties of TDEs, Taeho Ryu, Ho-Sang Chan, and Sasha Tchekhovskoy for the interesting discussions about numerical simulations, and Chris White for his help with Athena.  
    This work was supported by a grant from the Simons Foundation (00001470, AL, PA). OB was supported by an ISF grant 2067/22, a BSF grant 2018312, and an NSF-BSF grant 2020747.
\end{acknowledgments}

%
\begin{contribution}

PA has developed the version of the code and run the simulations. Data analysis and discussion contains equal contributions from all the co-authors. 


\end{contribution}

\software{\athena\ \citep{athena, athenapp}, NumPy \citep{harris2020array}, SciPy \citep{2020SciPy-NMeth}, Matplotlib \citep{matplotlib}, Plotly (\url{https://plotly.com/python/}) }


\appendix

\section{Parabolic motion}\label{sec:app:parabolic}

In polar coordinates, the shape of a parabola is set by the equation
\begin{equation}
r_*(\varphi) = \frac{2r_{\rm p}}{1+\cos\left( \varphi-\varphi_0\right)},
\end{equation}
where $r_{\rm p}$ is the pericenter distance, $\varphi_0$ sets the orientation of the orbit, and $\nu = \varphi-\varphi_0$ is the true anomaly of the orbit \citep{battin1987introduction,bate2020fundamentals}. 
Time anomaly may be converted to time using Barker's equation
\begin{equation}
\displaystyle t-t_0 = \sqrt{\frac{2r_{\rm p}^3}{GM}} \left[ D + \frac{1}{3}D^3\right],
\end{equation}
where 
\begin{equation}
D = \tan \frac{\nu}{2},
\end{equation}
and $t_0$ is the time of pericenter passage. 
Conversion from time to true anomaly may be done as
\begin{equation}
\displaystyle \nu = 2 \arctan \left[    
2\sinh \left(
\frac{1}{3} \asinh \frac{3m}{2}
\right)
\right],
\end{equation}
where $m$ is mean anomaly, 
\begin{equation}
\displaystyle m = \sqrt{\frac{GM}{2r_{\rm p}^3}} \left( t-t_0\right).
\end{equation}
Tangent to the parabolic trajectory is set by the vector
\begin{equation}\label{E:parabolic:tangent}
    \vector{T} = \frac{\left( \diff X/\diff \nu, \ \diff Y/ \diff\nu\right)}{\sqrt{(\diff X/ \diff \nu)^2 + ( \diff Y/\diff \nu)^2}} 
    = \frac{1}{\sqrt{2\left( 1+\cos \nu\right)}} \left( -\sin \nu , \ 1+\cos \nu\right),
\end{equation}
where $X$ and $Y$ are Cartesian coordinates in the equatorial plane.

\section{Polytropic solutions for the stellar structure}\label{sec:app:Emden}

Polytropic mass and pressure distribution is a natural guess for the initial conditions. The index of the polytrope is not in general related to the equation of state, as the star is likely to have initially some radial profile of entropy. Constant entropy suggests a polytropic index of $n=3/2$, that is a reasonable approximation for a fully convective stars. Radiative stellar interiors are closer to $n=3$ \citep{1967aits.book.....C}.
However, most of the solutions of the Lane-Emden equation are finite, meaning that density and pressure are exactly zero outside certain radius. Adding ambient medium to such a solution means to break the hydrostatic equilibrium in the initial conditions. 

Hence, we choose the only infinite finite-mass analytic solution of Lane-Emden equation, namely the $n=5$ polytrope. [ref] 
For this solution, 
\begin{equation}
\rho_*(R) = \rho_{\rm c} \theta^n(R/\alpha),
\end{equation}
\begin{equation}
P_*(R) = P_{\rm c} \theta^{n+1}(R/\alpha),
\end{equation}
where
\begin{equation}
\alpha = \sqrt{\frac{n+1}{4\pi G} \frac{P_{\rm c}}{\rho_{\rm c}^2}},
\end{equation}
and
\begin{equation}
    \theta(\xi) = \frac{1}{\sqrt{1+\frac{1}{3} \xi^2}}.
\end{equation}
The mass of the star may be found by integrating the density
\begin{equation}
M_* = 4\pi \int_0^{+\infty} R^2 \rho_*(R) \diff R = 4\pi \sqrt{3} \alpha^3 \rho_{\rm c}.
\end{equation}
Cumulative mass within some radius $R$
\begin{equation}
    M(R) = M_* \frac{R^3}{\left( R^2 + 3\alpha^2\right)^{3/2}}.
\end{equation}
The region $R < \alpha$ contains only $1/8$ of the total mass. 
For $\alpha=6\Rsun$, the stellar interior $R < 10\Rsun$ contains about $M_*/3 = 1\Msun$. The cut-off radius we use $R = 13.56\Rsun$ encloses about $0.5M_*$.
If the mass and radius $\alpha$ are set, the central density and pressure may be found as
\begin{equation}
\rho_{\rm c} = \frac{M}{4\pi \sqrt{3} \alpha^3},
\end{equation}
\begin{equation}
P_{\rm c} = \frac{4\pi G}{n+1} \alpha^2 \rho_{\rm c}^2 = \frac{G}{12\pi (n+1)} \frac{M^2}{\alpha^4}.
\end{equation}
Setting density and pressure cut-offs modifies the pressure and density profiles by a cut-off multiplier $C\left(\frac{R-R_{\rm CO}}{\Delta R_{\rm CO}}\right)$, where 
\begin{equation}
    C(x) = \frac{1}{2} \left( 1- \tanh{x} \right) =  \frac{1}{1+\e^{2x}}.
\end{equation}
Outside $r_{\rm CO}$, hydrostatic equilibrium is no more valid, and the medium is out of balance by a factor corresponding to the local pressure. 
In the initial conditions of our simulation, each cell is ascribed with the density equal to the maximum of $\rho_{\rm bgd}$ and $\rho_*(R) C\left(\frac{R-R_{\rm CO}}{\Delta R_{\rm CO}}\right)$, and pressure equal to the maximum of $P_{\rm bgd}$ and $P_*(R) C\left(\frac{R-R_{\rm CO}}{\Delta R_{\rm CO}}\right)$. 

\section{The impact of the ambient medium}\label{sec:app:dyne}

The density contrast between a star filling its Roche lobe ($\sim 10^{-3}-1\gcmc$) and the ambient medium around a quiescent BH is likely between ten and twenty orders of magnitude (see, for instance, \citealt[section 3.6]{kochanek94}). 
Reproducing such a density contrast is challenging for numerical simulations, hence we are using larger background densities, which is acceptable as long as the dynamic effect of the ambient matter is negligible. 
A reasonable estimate for the background density is set by the condition that the debris cloud is interacting with a mass smaller then its own. 
Taking $a_{\rm min}^3$ as an estimate for the volume being swept by the debris, we get a restriction
\begin{equation}
    \rho_{\rm amb} \lesssim \left( \frac{M_*}{M_{\rm BH}} \right)^2 M_* R_*^{-3}.
\end{equation}
For an $M_{\rm BH} = 3\times 10^3\Msun$, $\rho_{\rm amb} \sim 10^{-9}\gcmc$. 
To ensure that the negligible dynamical impact of the ambient gas, we apply a mass boost to the stellar material (increase the stellar density by a factor of $1000$) during the remapping procedure at $t\simeq 10$d, when the dimensions of the debris cloud are $\sim 1000 \times 100 \times 20\Rsun$. 

\section{Stellar magnetic field structure}\label{sec:app:fields}

Magnetic field is assumed to be non-zero in a range of radii, from $R_1=2\Rsun$ to $R_2=8\Rsun$. 
Two field topologies where considered: toroidal and poloidal. In both cases, the maximal magnetic field strength is normalized by the condition $\min \beta_{\rm m} = 1000$. In this section, we omit the normalization factors that are overridden by this renormalization. 

\subsection{Toroidal field}

Toroidal field is assumed uniform within a torus defined by the condition
\begin{equation}
 (r-R_0)^2+z^2 = R_{\rm m}^2, 
\end{equation}
where $R_0  = (R_1+R_2)/2$ and $R_{\rm m} = (R_2-R_1)/2$, and $r$ is cylindrical radius.
Vector potential was chosen to have a single component $A_z = A_z(r, z)$. 
\begin{equation}
A_z (r, z) = 
\left\{
\begin{array}{lc}
0 & \mbox{if $|z|\geq R_{\rm m}$}\\
0 &  \mbox{if $|z|<R_{\rm m}$, $r\leq R_{-}(z)$}\\
r-R_{-}(z) &  \mbox{if $|z|<R_{\rm m}$, $R_{-}(z) < r < R_{+}(z)$}\\
R_{+}(z)-R_{-}(z) & \mbox{if $|z|<R_{\rm m}$, $r\geq R_{+}(z)$},\\
\end{array}
\right. 
\end{equation}
where 
\begin{equation}
R_\pm(z) = R_0 \pm \sqrt{R_{\rm m}^2-z^2}
\end{equation}
are the inner and the outer surfaces of the torus. 
As
\begin{equation}
B_\varphi = - \pardir{r}{A_z},
\end{equation}
this vector potential encodes a uniform toroidal field within the torus. 

\subsection{Poloidal field}

To set a topologically dipolar poloidal magnetic field in a range of spherical radii from $R_1$ to $R_2$, it is sufficient to set 
\begin{equation}
  \displaystyle  A_\varphi = \left\{ 
    \begin{array}{cr}
   \displaystyle \frac{\sin\theta}{R} \left( R_2 -  R\right) \left( R-R_1\right) & \mbox{ if }R_1 < R < R_2,\\
   \displaystyle 0 & \mbox{otherwise}.
    \end{array}
    \right.
\end{equation}
Angular dependence ensures a dipole-like behavior with the radial magnetic field changing sign between the hemispheres. Polar angle component changes sign with the radial coordinate in the middle of the magnetized region ($R = R_0 = (R_1+R_2)/2$). 
The magnetic field component in this case are
\begin{equation}
    B_r = 2\frac{\left(R_2 - R\right)\left( R -R_1\right)}{R^2} \cos \theta
\end{equation}
and 
\begin{equation}
    B_\theta = \frac{2R - R_1 - R_2}{R} \sin\theta.
\end{equation}

\section{Remapping magnetic fields}\label{sec:app:avec}

Linear interpolation of the field components does not conserve the divergence of the field and produces $\div \vector{B} \neq 0$ on the new grid.
This is known to lead to catastrophic results, especially if the magnetic field lines are moving across the grid (see for instance \citealt{1980JCoPh..35..426B}).
A natural way to ensure zero divergence in the interpolated data is by calculating the magnetic field as $\vector{B} = \nabla \times \vector{A}$, where $\vector{A}$ is vector potential interpolated onto the new grid.
To calculate the vector potential on the new grid, we first calculated the magnetic field on a regular intermediate grid. 
Then, the field was used to calculate the current density as (assuming a Heaviside-Lorentz system with $c=1$)
\begin{equation}
    \vector{j} = \nabla \times \vector{B},
\end{equation}
and then solved Poisson equations separately for the vector potential components
\begin{equation}
    \nabla^2 \vector{A} = -\vector{j}.
\end{equation}
The solution of the Poisson equations were found using the relaxation scheme
\begin{equation}
\begin{array}{l}
 \displaystyle   \vector{A}^{n+1}_{ijk} = \frac{1}{6} \left( \vector{A}^{n}_{i-1, j, k} +  \vector{A}^{n}_{i+1, j, k} \right. \\
\displaystyle \left.    \qquad{} +\vector{A}^{n}_{i, j-1, k} +  \vector{A}^{n}_{i, j+1, k} + \vector{A}^n_{i, j, k-1} + \vector{A}^n_{i, j, k+1} + \vector{j}_{ijk} \Delta x^2\right),
    \end{array}
\end{equation}
where the upper index refers to the number of the iteration, and the lower indices to the coordinates of the cell. 
Both the currents and the vector potential components were defined on the cell edges in a way typical for a staggered grid \citep{1988ApJ...332..659E}.
Vector potential was restored on a uniform grid with a resolution that allows a good coverage of the stellar debris cloud. 
As a convergence criterion, we used $\max\left| \vector{A}^{n+1}-\vector{A}^n\right| < 10^{-4} \max \left|\vector{A}^n\right|$.

Each magnetic field component on the new grid was calculated in the center of the corresponding (orthogonal to the field direction) face using the definition of curl. 
Calculating the field involves taking a linear integral over the edges of the face.
For instance, for the $x$ component of the field,
\begin{equation}\label{E:Bxexample}
    \begin{array}{l}
 \displaystyle   \left(B_x\right)_{i+1/2, j, k} = \frac{1}{\Delta y \Delta z} \left( 
    \int_{i+1/2, j, k-1/2} A_y \diff y + \int_{i+1/2, j+1/2, k} A_z \diff z \right. \\
\displaystyle  \left. \qquad{}  - \int_{i+1/2, j, k+1/2} A_y \diff y - \int_{i+1/2, j-1/2, k} A_z \diff z     \right),
    \end{array}
\end{equation}
where the indices of the integrals refer to the central point of the relevant edge (if $i$, $j$, and $k$ set the center of a cell), and $\Delta x$, $\Delta y$, and $\Delta z$ are the grid cell sizes along the coordinates. 
Other field components were calculated in a similar way. 
To make the integration consistent among different refinement levels, we approximated each integral by a sum $N =2^{l_{\rm max}-l}$ points, where $l$ is the refinement level, and $l_{\rm max}$ is the maximal refinement level of the new grid. 
For the first integral in Eq.~(\ref{E:Bxexample}), 
\begin{equation}
    \int A_y \diff y \simeq \frac{\Delta y}{N} \sum_{m = 0..N-1} A_y\left( x_{i+1/2}, y_j + \frac{m+1/2}{N} \Delta y , z_{k-1/2} \right).
\end{equation}
For each point, the vector potential component was found by trilinear interpolation from the intermediate grid.

\bibliography{bibfile}{}
\bibliographystyle{aasjournalv7}



\end{document}